\pdfoutput=1
\newcommand*{\ATLASLATEXPATH}{latex/}
\documentclass[cernpreprint,texlive=2016,txfonts,UKenglish]{latex/atlasdoc}

\usepackage{\ATLASLATEXPATH atlaspackage}
\usepackage{\ATLASLATEXPATH atlasbiblatex}

\usepackage{\ATLASLATEXPATH atlasphysics}

\graphicspath{{logos/}{figures/}}

\usepackage{Backup_FourJetResonance-defs}
\usepackage{booktabs}
\usepackage{pdflscape}
\usepackage{multirow}
\usepackage{float}
\usepackage{graphicx}
\DeclareSIUnit[number-unit-product={}]{\percent}{\%}

\AtlasTitle{A search for pair-produced resonances in four-jet final states at $\sqrt{s}=13$ \TeV{} with the ATLAS detector}

\AtlasRefCode{SUSY-2016-09}

 \AtlasNote{SUSY-2016-09}

\PreprintIdNumber{CERN-EP-2017-183}

 \AtlasDate{\today}

 \AtlasJournalRef{Eur. Phys. J. C 78 (2018) 250}
 \AtlasDOI{DOI:10.1140/epjc/s10052-018-5693-4}

\AtlasAbstract{%
A search for  massive coloured resonances which are pair-produced and decay into two jets is presented.
The analysis uses $36.7$~\ifb{} of $\sqrt{s}$ = 13 TeV $pp$ collision data recorded by the ATLAS experiment at the LHC in 2015 and 2016.
No significant deviation from the background prediction is observed.
Results are interpreted in a SUSY simplified model where the lightest supersymmetric particle is the top squark, $\tilde{t}$, 
which  decays promptly into two quarks through  $R$-parity-violating couplings.
Top squarks with masses in the range $100~\GeV<m_{\tilde{t}}<410$~\GeV{}   are excluded at 95\% confidence level.
If the decay is into a $b$-quark and a light quark, a dedicated selection requiring two $b$-tags 
is used to exclude masses in the ranges $100~\GeV<m_{\tilde{t}}<470$~\GeV{} and $480~\GeV<m_{\tilde{t}}<610$~\GeV{}.
Additional limits are set on the pair-production of massive colour-octet resonances.
}

\hypersetup{pdftitle={ATLAS draft},pdfauthor={The ATLAS Collaboration}}

\begin{document}

\maketitle

\clearpage

\section{Introduction}
\label{sec:introduction}
Massive coloured  particles decaying into quarks and gluons are predicted in several extensions of the Standard Model (SM).
At hadron colliders, the search for new phenomena in fully hadronic final states, without missing transverse momentum,
is experimentally challenging due to the very large SM multijet production cross-section.
This paper describes a search for pair-produced  particles each decaying into two jets using 36.7~\ifb{} of $\sqrt{s}=13$~\TeV{} 
proton--proton ($pp$) collision data recorded in 2015 and 2016 by the ATLAS experiment at the Large Hadron Collider (LHC).

Supersymmetry (SUSY) \cite{Golfand:1971iw,Golfand2,Volkov:1973ix,Wess:1974tw,Wess:1974jb,Ferrara:1974pu,Salam:1974ig} is
a generalisation of the Poincar\'{e} symmetry group that  relates fermionic and bosonic degrees of freedom.
In the generic superpotential,  Yukawa couplings  can lead to baryon- and lepton-number violation:
\begin{equation*}\label{eq:rpv}
\mathcal{W}_{\textrm {RPV}} =   \lambda_{ijk}L_iL_j\overline{E}_k+\lambda'_{ijk}L_i Q_j\overline{D}_k+\lambda''_{ijk}\overline{U}_i\overline{D}_j\overline{D}_k + \kappa_iL_iH_u,
\end{equation*}
where $i$, $j$, and $k$ are quark and lepton generation indices.
The $L_i$ and $Q_i$ represent the lepton and quark $\mathrm{SU}(2)_\mathrm{L}$ doublet superfields
and $H_u$ the Higgs superfield that couples to up-type quarks.
The $\bar{E}_i$, $\bar{D}_i$, and $\bar{U}_i$  are the lepton, down-type quark
and up-type quark $\mathrm{SU}(2)_\mathrm{L}$ singlet superfields, respectively.
For each term the couplings are  $\lambda$, $\lambda'$, $\lambda''$, as well as $\kappa$ which is  a dimensional mass parameter. 
The $\lambda$ and $\lambda''$ couplings are antisymmetric in the exchange of $i\to j$ and $j\to k$, respectively.
While these terms in many scenarios are  removed by imposing an additional $Z_2$ symmetry ($R$-parity) \cite{Farrar:1978xj},
the possibility that at least some of these $R$-parity-violating (RPV) couplings
are not zero is  not ruled out experimentally~\cite{Barbier:2004ez,Dreiner:1997uz}.
This family of models leads to unique collider signatures which can escape conventional searches for $R$-parity-conserving SUSY.

Naturalness arguments~\cite{Barbieri:1987fn,deCarlos:1993yy}
suggest that higgsinos and top squarks\footnote{ The superpartners of the left- and right-handed top quarks, $\tilde{t}_\mathrm{L}$ and $\tilde{t}_\mathrm{R}$,
  mix to form the two mass eigenstates $\tilde{t}_1$ and $\tilde{t}_2$, where $\tilde{t}_1$ is the lighter one.
  This analysis considers only the production of the $\tilde{t}_1$, which hereafter is simply referred to as $\tilde{t}$.}
(stops) should be light,  with masses below a \TeV~\cite{Inoue:1982pi,Ellis:1983ed}.
Third-generation squarks in $R$-parity-conserving scenarios, and top squarks in particular,
have been the subject of a thorough programme of searches
at the LHC~\cite{SUSY-2014-07,SUSY-2015-02,SUSY-2016-15,CMS-SUS-14-015,CMS-SUS-13-023,CMS-SUS-14-006,CMS-SUS-16-008,CMS-SUS-16-009}.

If the top squark decays through RPV couplings, however,
the existing bounds on its mass can be significantly relaxed~\cite{Evans:2012bf,Brust:2012uf,Franceschini:2012za,Bai:2013xla}.
Indirect experimental constraints \cite{Allanach:1999ic} on the
sizes of each of the  $\lambda''$ couplings
are primarily valid for low squark mass and for first- and second-generation couplings.

This search targets a model where the top squark  is the lightest supersymmetric particle
and decays through  baryon-number-violating RPV $\lambda''$ couplings, $\tilde{t}\to \bar{q}_j\bar{q}_k$.
The couplings are assumed to be sufficiently large for the decays to be prompt,
but small enough to neglect the single-top-squark resonant production through RPV couplings.
Top squarks are then produced through strong interactions with cross-sections that do not depend on the specific assumptions in the SUSY model.
For two specific choices of couplings, the process considered is schematically depicted in Figure~\ref{fig:feynman-diagrams}.

In  models with extended SUSY, colour-octet states
can arise as scalar partners of a Dirac gluino~\cite{Kribs:2007ac,Choi:2008ub,Choi:2008pi,Plehn:2008ae}.
These scalar gluons (or sgluons) are mostly produced in pairs, and  decay into two quarks or two gluons.

Massive colour octet-resonances, generically referred to as colorons ($\rho$)~\cite{Kilic:2008pm,Alves:2011wf}
are predicted in a wide range of other theories, including axigluon~\cite{Frampton:1987dn, Bagger:1987fz} and topcolor \cite{Hill:1991at},
in vector-like confinement models \cite{Kilic:2009mi,Schumann:2011ji} and as Kaluza--Klein excitations of the gluons \cite{Lillie:2007yh,Burdman:2006gy}.
Colorons can be pair-produced and decay into two jets, a scenario which leads to a four-jet final state.

Constraints on top squarks decaying through $\lambda''$ couplings  
were first set by the ALEPH experiment at LEP~\cite{Heister:2002jc},
excluding at 95\% confidence level (CL) masses below 80~\GeV.
The CDF experiment at the Tevatron~\cite{Aaltonen:2013hya},
increased these limits to 100~\GeV.
Searches for pair-produced resonances in hadronic final states
were performed at the LHC  at 7~\TeV{} and 8~\TeV{} of centre-of-mass energy 
by both the ATLAS  \cite{SUSY-2012-18,SUSY-2015-05}
and CMS experiments \cite{CMS-EXO-11-016,CMS-EXO-12-052}.
For  decays  including heavy-flavour jets in the final state, exclusion limits at 95\% CL
on the mass of the top squark in the ranges  $100\, \GeV \leq m_{\tilde{t}}\leq 320$~\GeV{}
and  $200\, \GeV \leq m_{\tilde{t}}\leq 385$~\GeV{} have been reported
by  ATLAS \cite{SUSY-2015-05} and CMS \cite{CMS-EXO-12-052}, respectively.

\begin{figure}
  \centering
  \subfloat[$\tilde{t}\tilde{t^*}\rightarrow (\bar{d}\bar{s})(ds)$]{\includegraphics[width=0.4\textwidth]{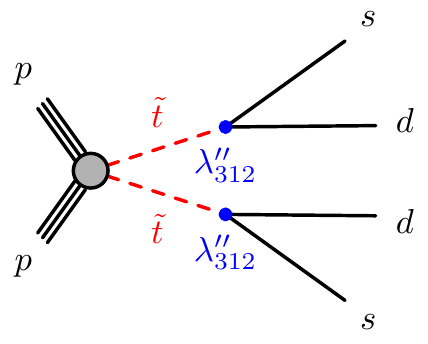}}
  \subfloat[$\tilde{t}\tilde{t^*}\rightarrow (\bar{b}\bar{s})(bs)$]{\includegraphics[width=0.4\textwidth]{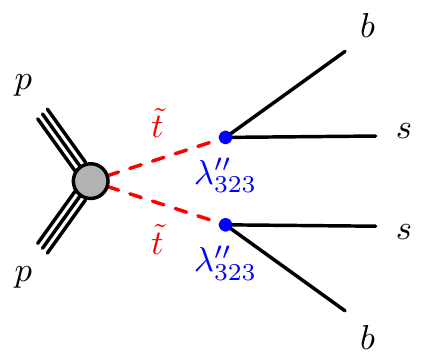}}\\
  \caption{\label{fig:feynman-diagrams} Diagrams depicting the direct pair-production of  top squarks through strong interactions,
   with decays into a $d$- and an $s$-quark (left) or into a $b$- and an $s$-quark (right) through the  $\lambda''$ $R$-parity-violating couplings,
   indicated by the blue dots.}
\end{figure}

\section{ATLAS detector}
The ATLAS detector~\cite{PERF-2007-01} is a multi-purpose particle physics detector with a forward-backward symmetric cylindrical geometry
and nearly $4\pi$ coverage in solid angle\footnote{ATLAS uses a right-handed coordinate system with its origin at the nominal interaction point in the centre of the detector.
  The positive $x$-axis is defined by the direction from the interaction point to the centre of the LHC ring,
  with the positive $y$-axis pointing upwards, while the beam direction defines the $z$-axis.
  Cylindrical coordinates ($r$, $\phi$) are used in the transverse plane, $\phi$ being the azimuthal angle around the $z$-axis.
  The pseudorapidity $\eta$ is defined in terms of the polar angle $\theta$ by $\eta = - \ln\tan(\theta/2)$.
  Rapidity is defined as $y = 0.5\cdot\ln[(E + p_z)/(E - p_z)]$ where $E$ denotes the energy and $p_z$ is the component of the momentum along the beam direction.}.
The inner tracking detector consists of pixel and silicon microstrip detectors covering the pseudorapidity region $|\eta| < 2.5$,
surrounded by a transition radiation tracker which provides electron identification in the region $|\eta| < 2.0$.
Starting in Run~2, a new inner pixel layer, the Insertable B-Layer (IBL)~\cite{Capeans:1291633,CapeansAddendum}, has been inserted at a mean sensor radius of 3.3 cm.
The inner detector is surrounded by a thin superconducting solenoid providing an 2 T axial magnetic field
and by a lead/liquid-argon (LAr) electromagnetic calorimeter covering $|\eta| < 3.2$.
A steel/scintillator-tile calorimeter provides hadronic coverage in the central pseudorapidity range ($|\eta| < 1.7$).
The endcap and forward regions ($1.5 < |\eta| < 4.9$) of the hadronic calorimeter are made of LAr active layers
with either copper or tungsten as the absorber material. An extensive muon spectrometer with an air-core toroidal magnet system surrounds the calorimeters.
Three layers of high-precision tracking chambers provide coverage in the range $|\eta| < 2.7$,
while dedicated fast chambers allow triggering in the region $|\eta| < 2.4$.
The ATLAS trigger system consists of a hardware-based level-1 trigger followed by a software-based high level trigger \cite{TRIG-2016-01}.

\section{Data sample}\label{sec:trigger}
The data used in this analysis were collected by the ATLAS detector in $pp$ collisions at $\sqrt{s} = 13$~\TeV{}
at the LHC using a minimum  proton bunch crossing interval of 25 ns during 2015 and 2016.
In this dataset the mean number of  $pp$ interactions per proton bunch crossing is about 23.
Events were recorded using a four-jet trigger with transverse momentum (\pt) thresholds of 100 GeV for each jet at the
high-level trigger, which is fully efficient after the analysis selection requirements are applied.
After requiring quality criteria for the beam, the data and the detector condition, 
the available dataset corresponds to an integrated luminosity of 36.7~\ifb{}
with an  uncertainty of $\pm2.1\%$ for the 2015 data and $\pm3.4\%$ for the 2016 data.
The uncertainty in the integrated luminosity is obtained from a  calibration
of the luminosity scale using a pair of beam-separation scans performed in August 2015 and June 2016,
following a methodology similar to that detailed in Ref.~\cite{DAPR-2013-01}.

\section{Simulated samples}\label{sec:models}

The dominant background from SM multijet production is estimated with a data-driven technique,
while Monte Carlo (MC) simulated events are used to estimate the contribution of the $t\bar{t}$ background,
to model the signals and to establish and validate the background estimation method.

The response of the detector was simulated~\cite{SOFT-2010-01}  using either a  GEANT4 simulation~\cite{Agostinelli:2002hh}
or a fast parameterised simulation~\cite{ATL-PHYS-PUB-2010-013} of the calorimeter response and GEANT4 for everything else.
To account for additional $pp$  interactions in the same and nearby bunch crossings (pile-up),
a set of minimum-bias interactions was generated using \textsc{Pythia} 8.186~\cite{Sjostrand:2007gs}  with the A2 set of parameters (tune)~\cite{ATL-PHYS-PUB-2012-003}
and the MSTW2008LO~\cite{Martin:2009iq,Sherstnev:2007nd} parton distribution function (PDF) set and was superimposed on the hard scattering events.
The EvtGen v1.2.0 program~\cite{EvtGen} was used to simulate properties of bottom and charm hadron decays for all samples.
Corrections were applied to the simulated events to account for differences between data and simulation
for the efficiency of identifying jets originating from the
fragmentation of $b$-quarks, together with the probability for mistagging light-flavour and charm-quark jets.

Background samples of multijet production were simulated with $2\to2$ matrix elements (ME)
at leading order (LO) using the \textsc{Pythia} 8.186 event generator.
The renormalisation and factorisation scales were set to the average \pt{} of the two leading jets.
The ATLAS A14 tune~\cite{ATL-PHYS-PUB-2014-021} of parton shower and multiple parton interaction
parameters  was used together with the NNPDF23LO PDF set~\cite{Ball:2012cx}.

Top-pair production events were simulated using the \textsc{Powheg-Box} v2 \cite{Alioli:2010xd} generator
with the CT10 PDF set. The top mass was set to 172.5 \GeV.
The $h_{\mathrm{damp}}$ parameter, which regulates the transverse momentum of the first extra gluon emission beyond the Born configuration
(and thus controls the transverse momentum of the $t\bar{t}$ system), was set to the mass of the top quark.
The parton shower, hadronisation, and  underlying event were simulated using \textsc{Pythia} 6.428 \cite{Sjostrand:2006za}
with the CTEQ6L1 PDF set and the corresponding Perugia 2012 tune (P2012) \cite{Skands:2010ak}.
The sample was normalised using the next-to-next-to-leading-order (NNLO) cross-section
including the resummation of soft gluon emission at next-to-next-to-logarithmic (NNLL) accuracy using \verb|Top++2.0| \cite{Czakon:2011xx}.

The search considers three benchmark signals: the pair production of top squarks, colorons and sgluons
with decays into two jets for each resonance.

Signal samples were generated using \textsc{MG5}\_a\textsc{MC@NLO}~\cite{Alwall:2014hca} v2.2.3 interfaced to \textsc{Pythia} 8.186 with
the A14 tune for the modelling of the parton shower, hadronisation and underlying event.
The ME calculation was performed at leading order and, for the top squark signal,
includes the emission of up to two additional partons.
The merging with the parton shower was done using the \mbox{CKKW-L}~\cite{Lonnblad:2012ix} prescription,
with a merging scale set to one quarter of the pair-produced resonance mass.
The PDF set used for the generation is NNPDF23LO.
For the top squark signal generation all the non-SM particle masses
were set to 5 \TeV{} except for the top squark mass ($m_{\tilde{t}}$) itself.
The top squark was decayed in \textsc{Pythia} 8
assuming a 100\% branching ratio into $\bar{b}\bar{s}$.
Its width is expected to be small, and negligible with respect to the detector resolution.
This set of samples is also used to interpret the analysis for the case where both top squarks decay into light quarks,
since the analysis  is not sensitive to the flavour content of the jets.
The top squark pair-production cross-sections were calculated at next-to-leading order (NLO) in the strong coupling constant,
adding the resummation of soft gluon emission at next-to-leading-logarithmic accuracy~\cite{Beenakker:1997ut,Beenakker:2010nq,Beenakker:2011fu}.
The nominal cross-section and its uncertainty were taken from an envelope of cross-section predictions
using different PDF sets and factorisation and renormalisation scales, as described in Ref.~\cite{Borschensky:2014cia}.
The coloron samples were generated with the model described in Ref.~\cite{Redi:2013eaa},
where the couplings of the vector colour octet to all particles except light quarks were set to zero.
The LO cross-sections from the event generator were used.
The coloron samples are also used to interpret the result in the context of sgluon pair-production,
where they are scaled to the sgluon cross-section computed at NLO with \textsc{MG5}\_a\textsc{MC@NLO}~\cite{GoncalvesNetto:2012nt,Degrande:2014sta}.
The sgluons are assumed to decay into two gluons, which in this analysis are not distinguished from quark-initiated jets. 

\section{Event reconstruction}
\label{sec:objectsreconstruction}

Candidate jets are reconstructed from three-dimensional topological energy clusters \cite{ATL-LARG-PUB-2008-002}
in the calorimeter using the anti-$k_t$ jet algorithm \cite{Cacciari:2008gp},
as implemented in the FastJet package~\cite{Cacciari:2011ma}, with a radius parameter of 0.4.      
Each topological cluster is calibrated to the electromagnetic energy scale prior to jet reconstruction.
The reconstructed jets are then calibrated to the particle level by the application of a jet energy scale (JES)
calibration derived from simulation and in situ corrections based on 13 \TeV{} data \cite{ATLAS-CONF-2013-004,ATLAS-CONF-2015-057,,ATL-PHYS-PUB-2015-015}.
The \verb|TightBad| cleaning quality criteria~\cite{ATLAS-CONF-2015-029} are imposed to identify jets arising from non-collision sources
or detector noise. Any event containing at least one jet failing quality requirements with $\pt{}>20$~\GeV{} is removed.

Jets containing $b$-hadrons ($b$-jets) are tagged by a multivariate algorithm (MV2c10) using information about the impact parameters
of inner detector tracks associated with the jet, the presence of displaced secondary vertices, 
and the reconstructed flight paths of $b$- and $c$-hadrons inside the jet \cite{PERF-2012-04}.
A working point with a 77\% efficiency, as determined in a simulated sample of $t\bar{t}$ events, is chosen.
The corresponding rejection factors against simulated jets originating from $c$-quarks
 and from light quarks or gluons are 4.5 and 130, respectively~\cite{ATL-PHYS-PUB-2016-012}.

\section{Event selection}\label{sec:eventselection}
Each event is required to have a reconstructed primary vertex 
with at least two associated tracks with $\pt > 400$ \MeV{} and a position consistent with the beamspot envelope.
If more than one such vertex is found, the vertex with the largest $\sum\pt^2$ of the associated tracks is chosen.

The final state under consideration consists of four jets forming two pairs
originating from a pair of equal-mass resonances.
After the trigger requirement, only events with at least four reconstructed jets
with $\pt > 120$~\GeV{} and $|\eta| < 2.4$ are retained in the analysis.

The analysis strategy exploits the case where the resonances are produced with a significant transverse momentum.
As a result  the decay products are expected to be close to each other.
Taking advantage of this property,
candidate resonances are constructed by pairing the four leading jets in the event.
Two jet pairs are identified by  the following quantity:
\begin{equation*}
\Delta R_{\text{min}} = \text{min}\left\{ \sum_{i=1,2} |\Delta R_{i} - 1| \right\},
\end{equation*}
where $\Delta R_{i}$ is the angular distance between the two jets for the $i^{\mathrm{th}}$ pair and the sum
is over the two pairs of dijets. The offset of $-1$ is chosen to maximise the signal efficiency for the
masses of interest while minimising the effects of soft jets from radiated gluons being recombined with
their parent jets in multijet topologies.

The above criteria define the analysis preselection.
Additional requirements are applied to further enhance the signal purity.
These are based on four discriminating variables
established from simulation studies and previous ATLAS searches~\cite{SUSY-2015-05,SUSY-2012-18,Schumann:2011ji}.

To reduce the non-resonant multijet background, for which the pairing efficiency is
expected to be poor, a quality criterion is applied to the pairing metric.
Resonances with higher masses are produced with a lower boost, and their decay products are less collimated.
To compensate for the larger (smaller) angular separation between the jets at high (low) mass 
this requirement is made dependent on the average reconstructed mass of the two resonance candidates in the event, \mavg.
The event is discarded if  the best combination of the four leading jets satisfies:
\begin{align*}
\Delta R_{\mathrm{min}} &>  -0.002 \cdot (m_{\mathrm{avg}}/\GeV-225) + 0.72  &  \mathrm{if}\; & m_{\mathrm{avg}} \leq 225~\GeV, \\
\Delta R_{\mathrm{min}} &>  +0.0013 \cdot (m_{\mathrm{avg}}/\GeV-225) + 0.72 & \mathrm{if}\; & m_{\mathrm{avg}} > 225~\GeV.
\end{align*}

After boosting the system formed by the two resonances into its centre-of-mass frame, the magnitude of the
cosine of the angle that either of them forms with the beamline is denoted as $|\cos(\theta^*)|$. Background jets from 
multijet production  frequently originate from $t$-channel gluon exchange and are preferentially produced
in the forward region, with $|\cos(\theta^*)|$ close to one. Jets originating from the signal are instead expected to
be more central and lead to small $|\cos(\theta^*)|$ values.

Since the two reconstructed resonances are expected to have equal mass, 
their mass difference is a powerful discriminant between signal and background.
The mass asymmetry ($\mathcal{A}$) is defined as:
\begin{equation*}
\mathcal{A} = \frac{|m_{1} - m_{2}|}{m_{1} + m_{2}}\, ,
\end{equation*}
where $m_{1}$ and $m_{2}$ are the invariant masses of the two reconstructed dijet pairs.
The value of $\mathcal{A}$ is expected to peak at zero for well-paired signal events and to have larger values for background events

\begin{figure}
  \centering
  \subfloat[]{\includegraphics[width=0.49\textwidth]{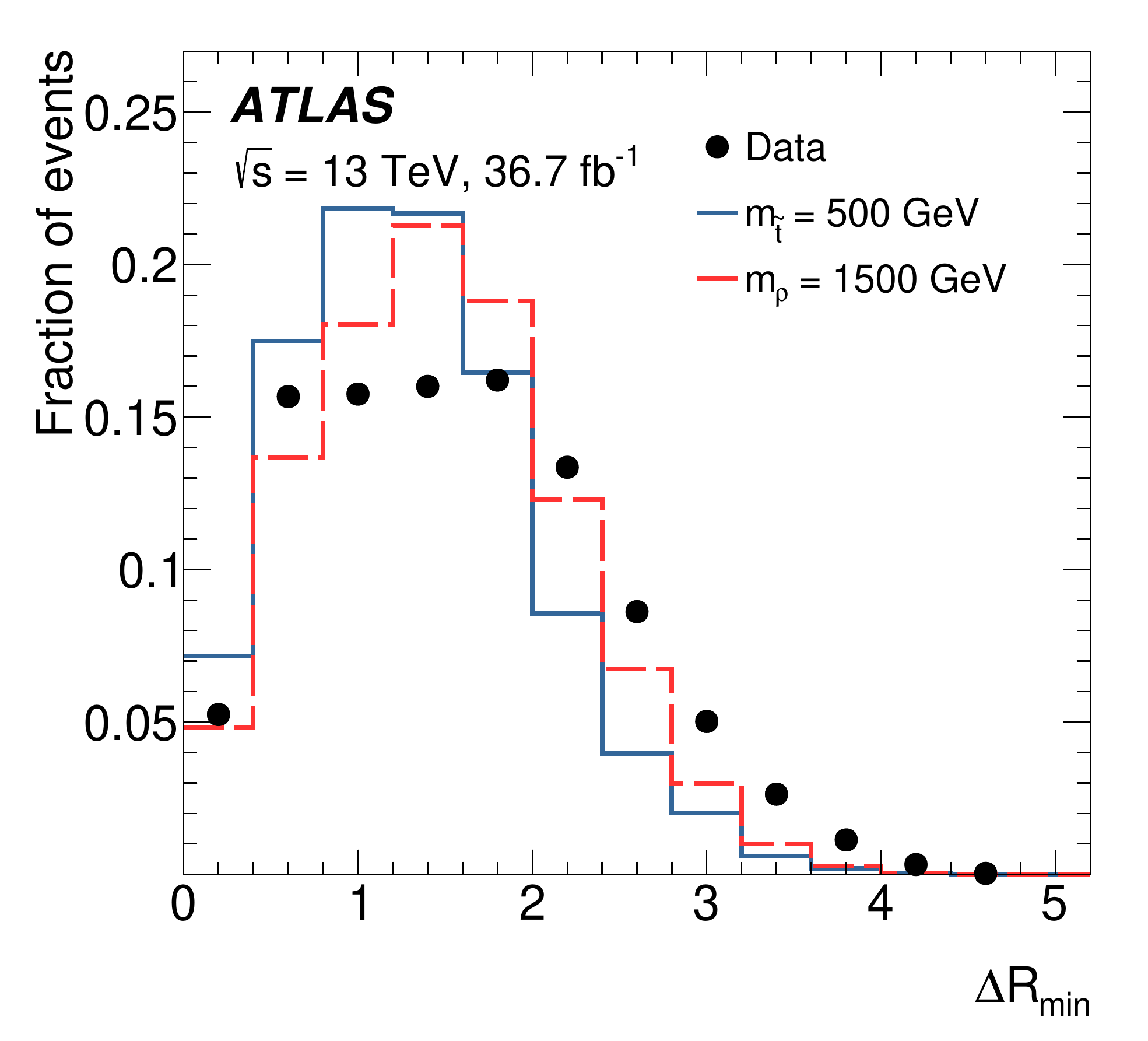}}
  \subfloat[]{\includegraphics[width=0.49\textwidth]{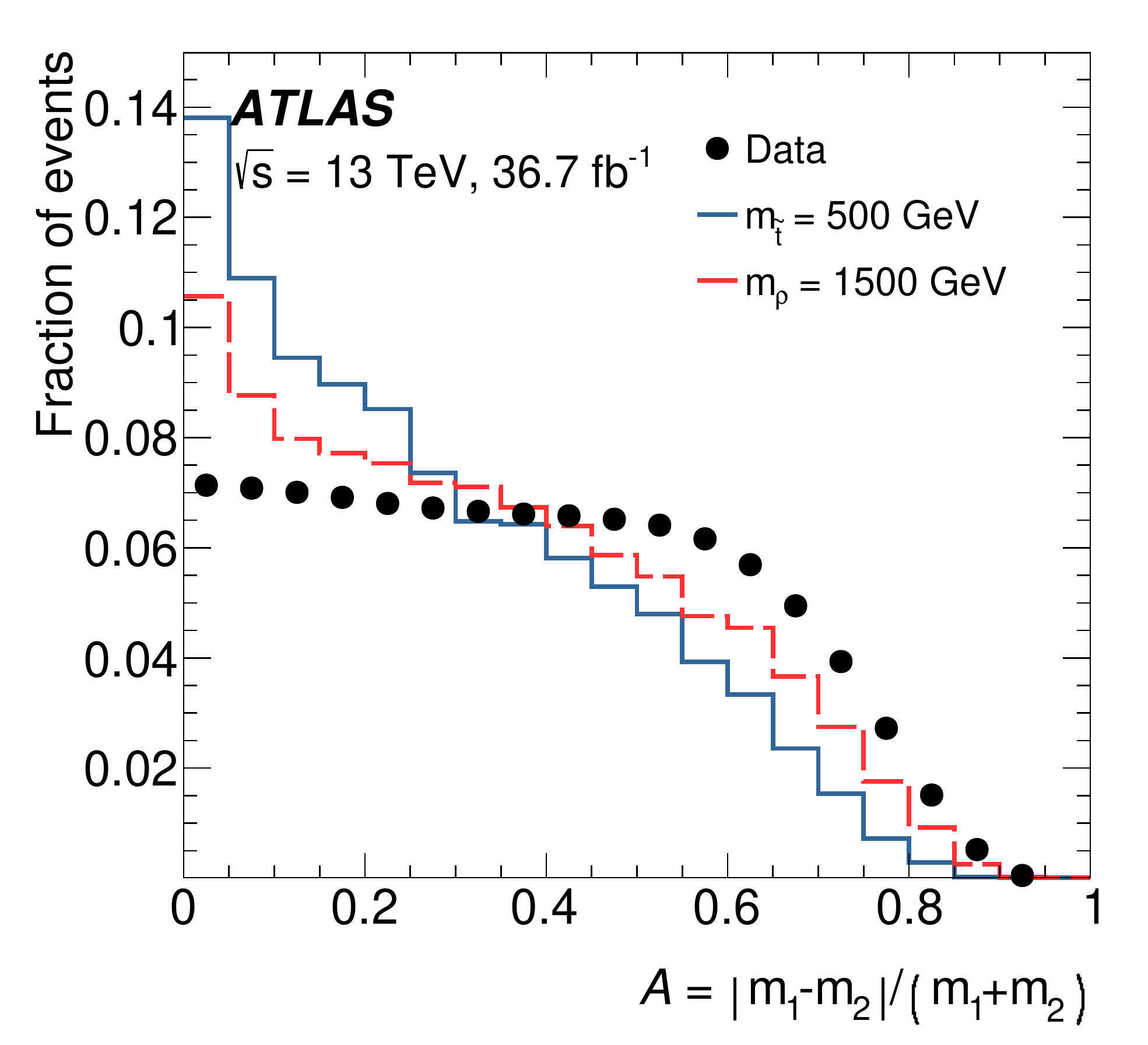}}\\
  \subfloat[]{\includegraphics[width=0.49\textwidth]{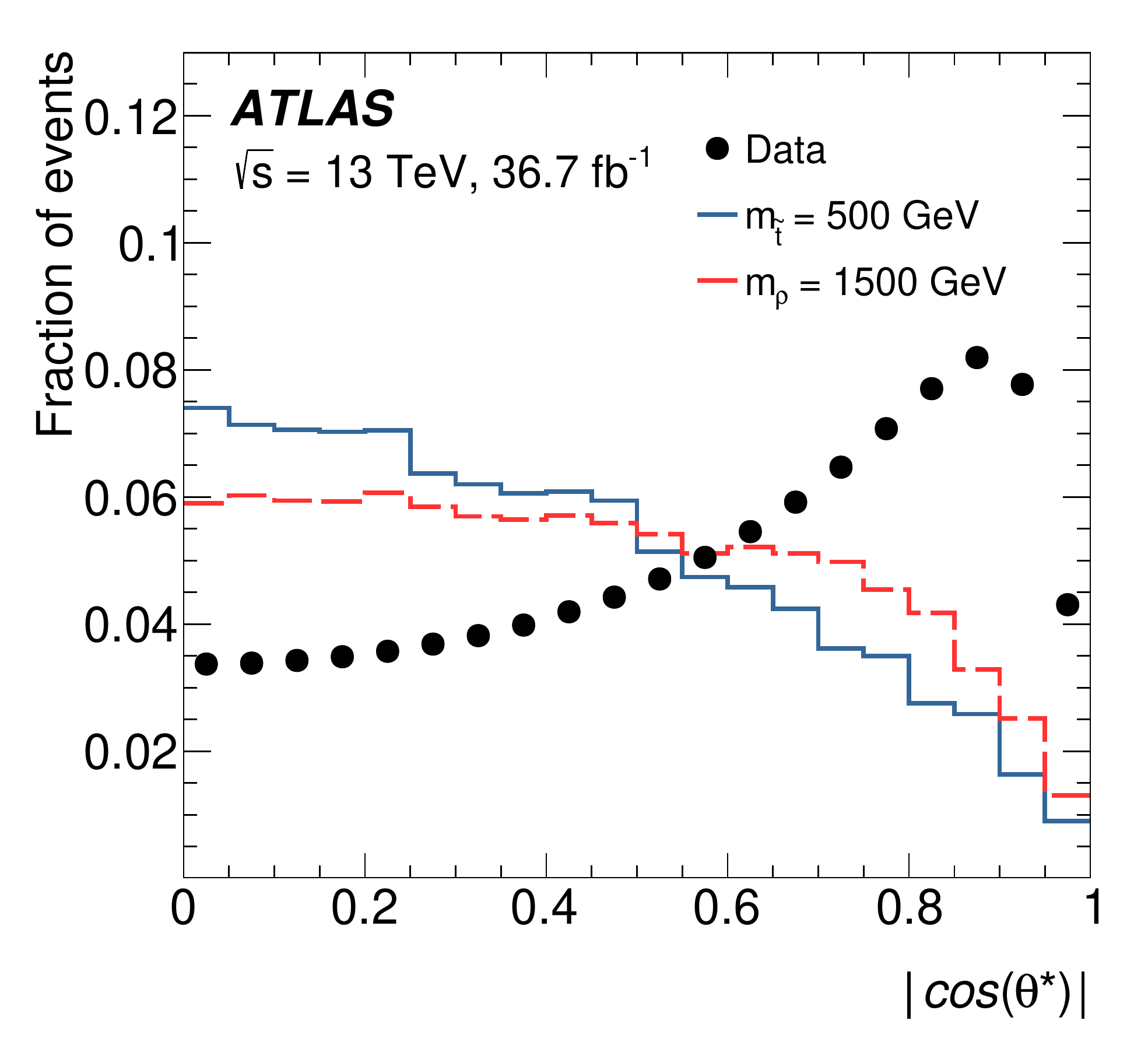}}
  \subfloat[]{\includegraphics[width=0.49\textwidth]{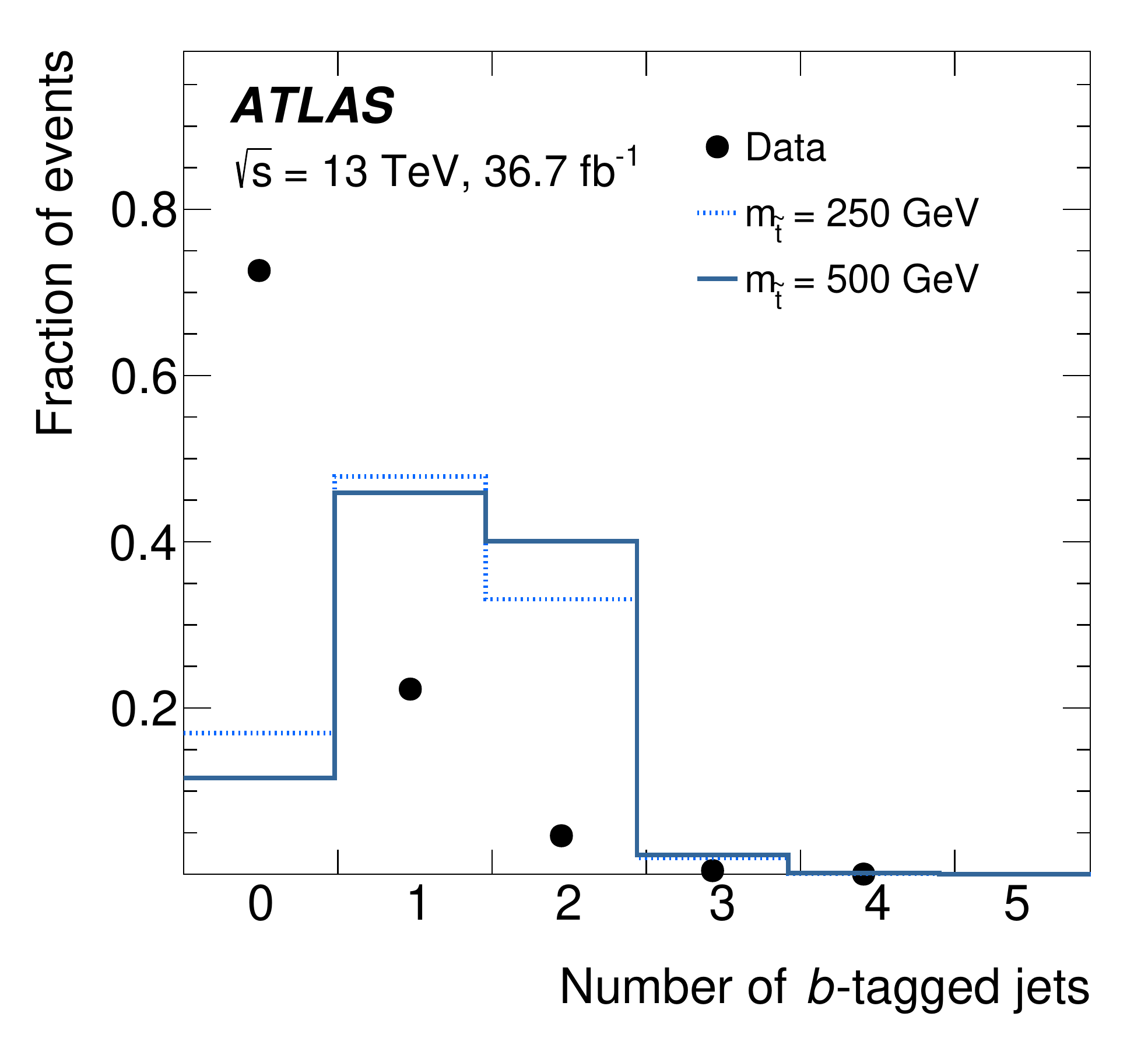}}
  \caption{\label{fig:variables} The distributions of the (a) smallest angular separation between the two jets in a pair ($\Delta R_{\textrm{min}}$),
    the (b) mass asymmetry ($\mathcal{A}$), the (c) pair production angle $|\cos(\theta^*)|$    and the (d) multiplicity of $b$-tagged jets.
    The observed data (black dots) are compared with the distributions expected from a top squark with a mass of 250~\GeV{} (solid blue line)
    or 500~\GeV{} (azure dotted line) and a coloron with a mass of 1500~\GeV{} (red dashed line).
    The distributions are normalised to unity and shown at preselection, after the requirement of four jets paired into two candidate resonances.}
\end{figure}

The distributions of $\Delta R_{\mathrm{min}}$, $\mathcal{A}$ and $|\cos(\theta^*)|$ after preselection
are shown for data, a top squark sample with a mass of $m_{\tilde{t}}=500$~\GeV{}
and a coloron sample with mass $m_{\rho}=1500$~\GeV{} in Figure~\ref{fig:variables}(a--c).
Because of the very small expected signal purity (below 2\%) before additional selection criteria are applied
 the data distributions can be viewed  as representative of the expected background. 
Two additional requirements, $\mathcal{A} < 0.05$ and $|\cos(\theta^*)|<0.3$, 
define the inclusive signal region (SR) selection, targeting resonance decays into light quark  or gluon jets.
The selections are determined in an optimisation procedure that maximises the expected signal significance.

When the dominant RPV couplings involve third-generation quarks ($\lambda^{''}_{3i3}$),
a $b$-quark is expected from each of the top squark decays.
A dedicated  $b$-tagged SR selection is used for this scenario.
On top of the requirements applied in the inclusive selection 
it requires at least two $b$-tagged jets to be present in the event,
which significantly reduces the multijet background.
The distribution of the number of $b$-tagged jets after pairing the four jets into candidate resonances 
is shown for data and two top squark signals with masses of 250 and 500~\GeV{} in Figure~\ref{fig:variables}(d).
An additional factor of about two in background reduction is gained
by requiring each of the two $b$-jets to be associated with a different reconstructed resonance.
This is particularly effective in reducing the contribution of $g\rightarrow b\bar{b}$ splittings, where the two $b$-jets are typically very collimated.

The final analysis discriminant is the average mass of the two reconstructed resonances:
\begin{equation*}
m_{\mathrm{avg}}=\frac12(m_1+m_2)\,.
\end{equation*}

A peak in \mavg{} at a mass of about that of the resonance is expected for the signal,
over a non-peaking background from multijet processes.
Figure~\ref{fig:signal} shows the expected \mavg{} distribution for signal samples with different masses.
For each mass hypothesis a counting experiment is performed in
a window of the \mavg{} variable optimised to maximise the expected signal significance.
The windows range from a 10~\GeV{} width for a 100~\GeV{} top squark to a 200~\GeV{} width for a 1500~\GeV{} coloron.
The mass window for the highest target mass considered, of 2000~\GeV{}, has no upper edge.
When the mass difference between two signal samples is smaller than 
the experimental resolution, their selected mass windows partially overlap.

\begin{figure}
  \centering
  \includegraphics[width=0.7\textwidth,height=0.6\textwidth]{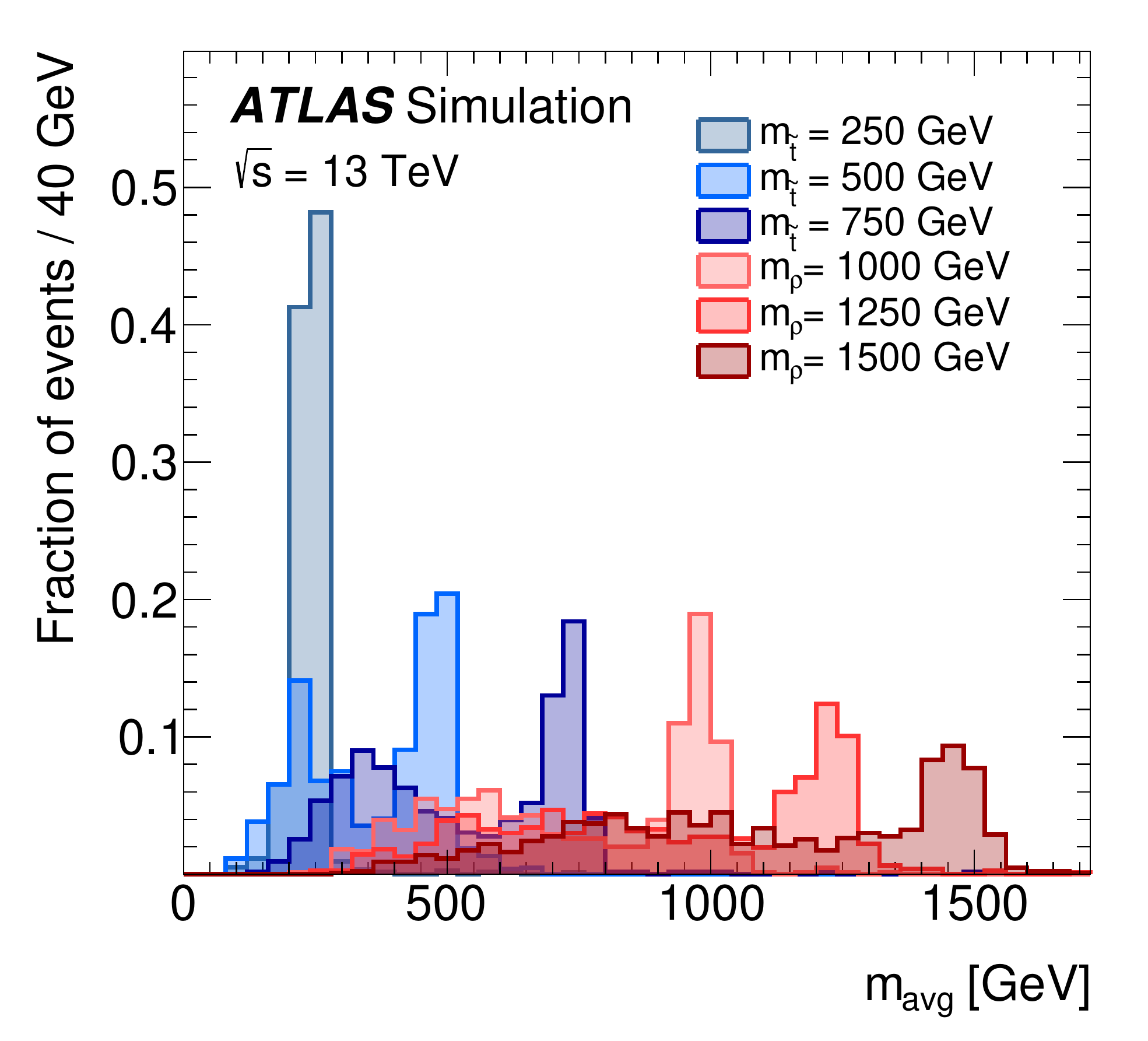}\label{subfig:signalmavg}
  \caption{\label{fig:signal}  Distribution of the average mass, \mavg, in the inclusive signal region 
    for simulated top squark signals with $m_{\tilde{t}}=250$~\GeV, 500~\GeV, and 750~\GeV{} 
    and coloron signals with $m_{\rho}=1000$~\GeV, 1250~\GeV, and 1500~\GeV. }
\end{figure}

\begin{figure}
  \centering
  \subfloat[]{\includegraphics[width=0.49\textwidth]{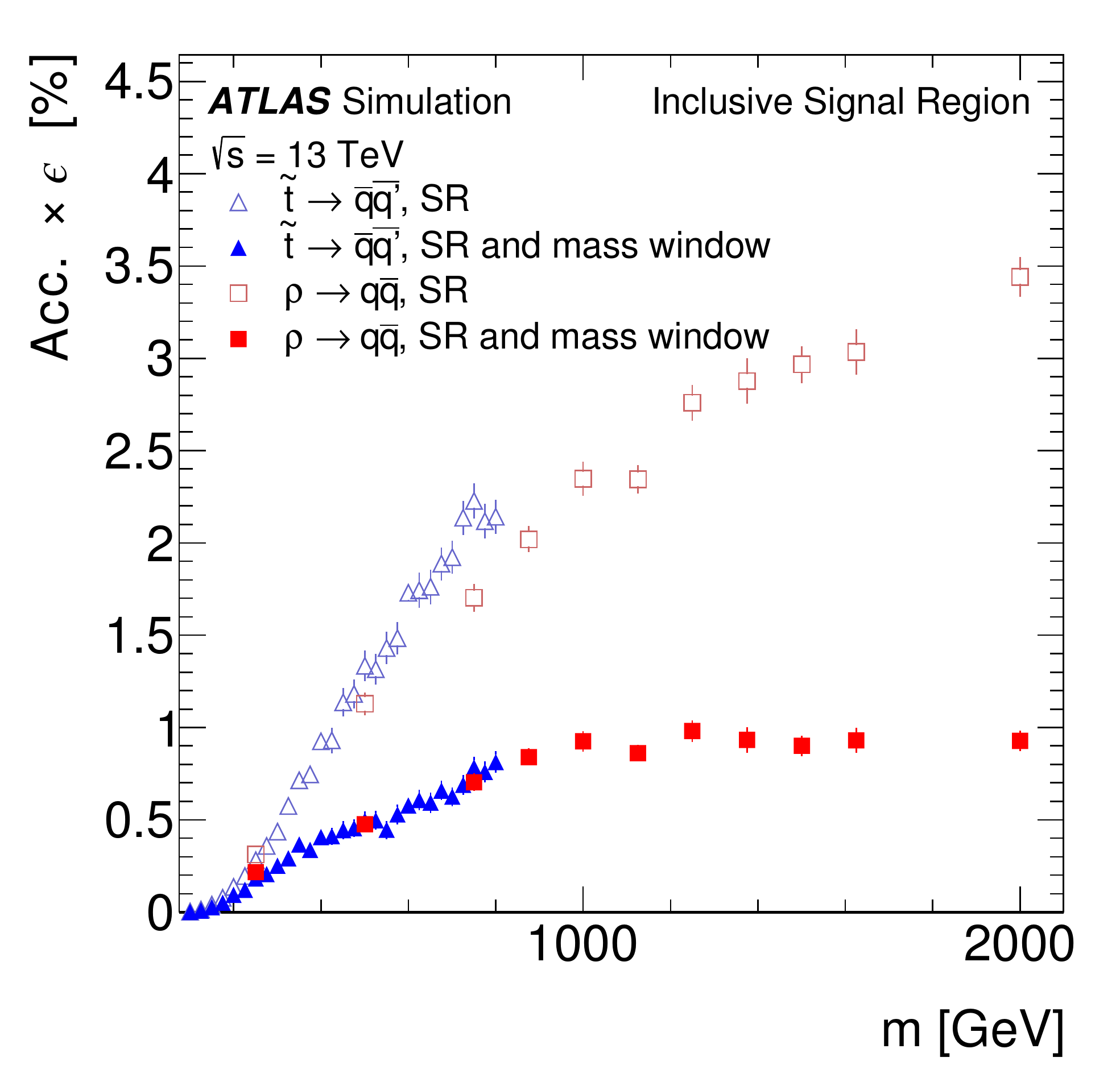}}
  \subfloat[]{\includegraphics[width=0.49\textwidth]{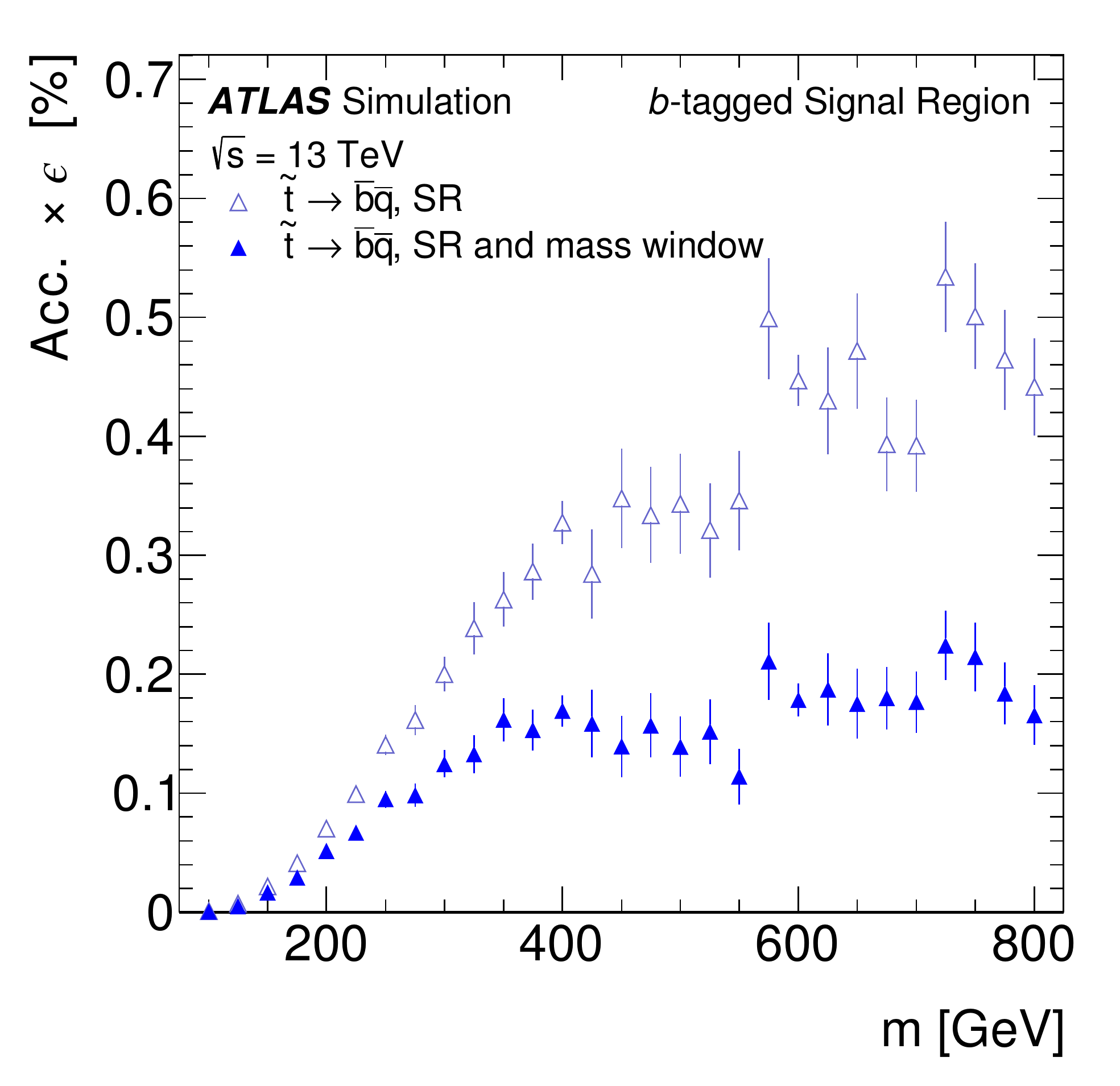}}
  \caption{\label{fig:acceptance} The acceptance times efficiency (Acc.$\times\epsilon$) of
    the (a) inclusive and (b) $b$-tagged  signal region selection  as a function of the resonance mass, $m$,
    before and after the mass window requirements are applied.
    Top squark signals are indicated by the blue triangles,  coloron by the red squares.
    The statistical uncertainties are indicated by  vertical bars.}
\end{figure}

\begin{table}\centering
  \caption{\label{tab:cutflow}  MC predictions of the number of signal events corresponding to 36.7 fb$^{-1}$ of data
    after applying each of the event selection requirements, except for the mass window.
    Top squark masses of $m_{\tilde{t}}=100$~\GeV{} and $m_{\tilde{t}}=500$~\GeV{}, and a coloron mass of 1500~\GeV{} are shown.
    The statistical uncertainty of the MC simulation is shown for each selection. }
  \sisetup{group-minimum-digits=5}
  \begin{tabular}{
      l 
      S[table-space-text-post = \si{\pm}]
      S[table-space-text-post = \si{\pm}]
      S[table-space-text-post = \si{\pm}]
    }
    \toprule
    Selection         &  {$m_{\tilde{t}}=100$~\GeV} & {$m_{\tilde{t}}=500$~\GeV} & {$m_{\rho}=1500$~\GeV} \\
    \midrule
    Total             & {$(558.0\pm0.6)\cdot 10^{5}$} & 19000\pm130     &   1710\pm10 \\
    Trigger           & 221900\pm420 & 11900\pm100       &   1650\pm10 \\
    $\Delta R_{\mathrm{min}}$  & 18910\pm120 & 2470\pm50    &   1050\pm5 \\
    \midrule
    {Inclusive selection}          & 1359\pm36         & 253\pm16   &   51\pm2 \\
    \midrule
    {$b$-tagged selection}            &   569\pm24         & 65\pm8     &   {--} \\
    \bottomrule
  \end{tabular}
\end{table}

The MC predictions of signal event yields in 36.7 fb$^{-1}$ of data are shown in Table~\ref{tab:cutflow}
after each different requirement of the event selection is applied.
The acceptance times efficiency of the inclusive and $b$-tagged signal region selections as a function of the signal mass
are shown before and after applying the \mavg{} mass window requirement in Figure~\ref{fig:acceptance}.
The acceptance of the signal region selections increases for large masses due to the four jets from the signal having a larger \pt.
However, as the jet pairing does not always correctly assign the resonance candidates for high masses,
the signal has a tail extending to low \mavg{} values, degrading the efficiency of the mass window selection.

\section{Background estimation}\label{sec:background}
The dominant background from multijet production  is estimated directly from data,
with a method that predicts both the normalisation and the shape of the \mavg{} distribution.
In the $b$-tagged selection, for \mavg{} below 200~\GeV, the  \ttbar contribution becomes significant,
and is estimated from simulation.

\begin{figure}[!htb]
  \centering
  \includegraphics[width=0.8\textwidth]{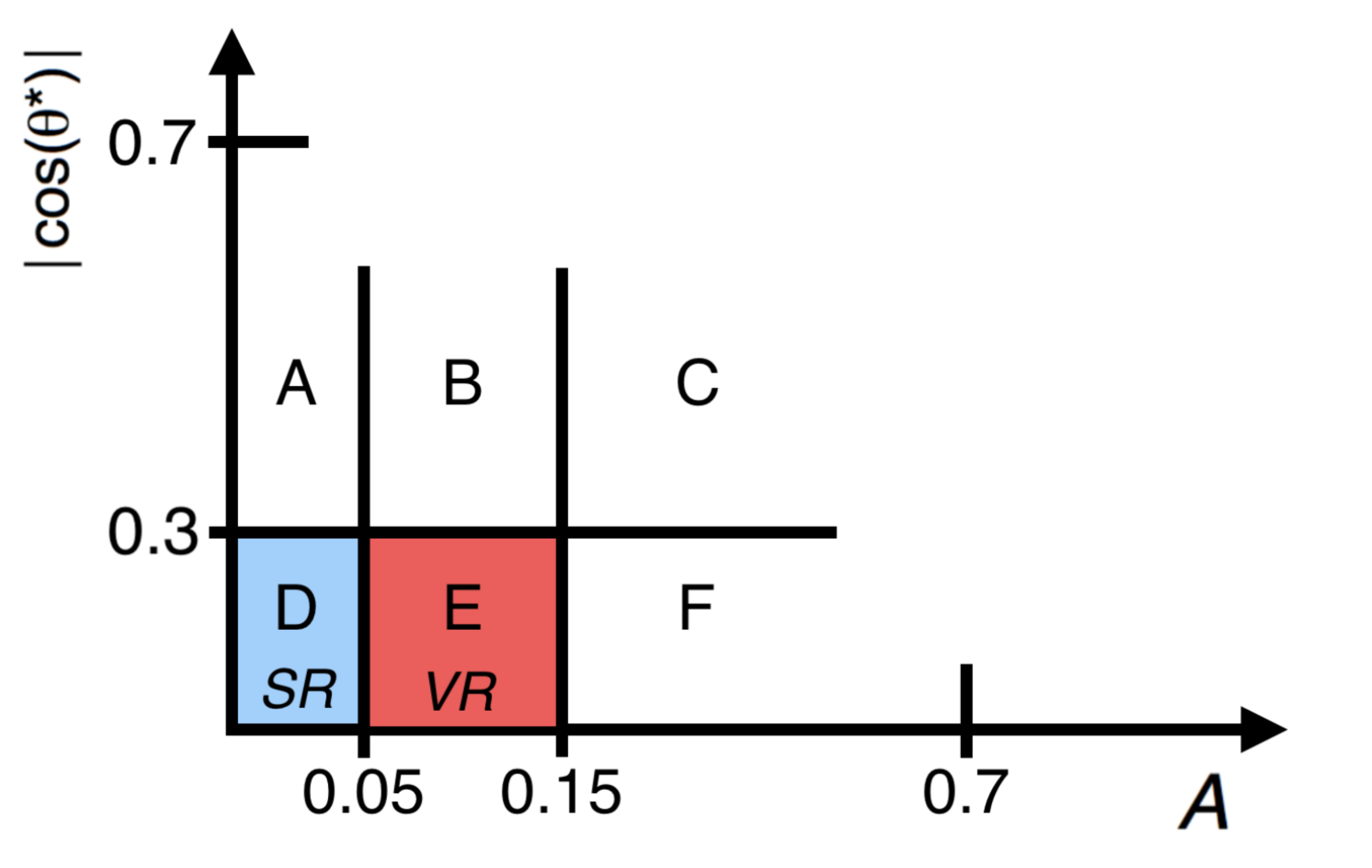}
    \caption{\label{fig:ABCDscheme} Definition of the control and validation regions in the  $\mathcal{A}$ and $|\cos(\theta^*)|$ plane
      used to estimate the multijet background.}
\end{figure}

For the inclusive selection, the \mavg{} distribution for the background is obtained from data.
For each \mavg{} bin the data sample is divided into four regions: one region where the signal region selection is applied (D)
and three background-dominated control regions (A, C and F).
The variables used to define the different regions, 
summarised in Figure~\ref{fig:ABCDscheme}, are $\mathcal{A}$ and $|\cos(\theta^*)|$.
Provided the two variables defining the regions are uncorrelated,
and signal leakage in the background-dominated regions can be neglected,
the amount of background in the region of interest D
can be predicted from the observed numbers of events in the control regions as $N_\mathrm{D} = N_\mathrm{A} \times N_\mathrm{F}/N_\mathrm{C}$.
The linear correlation between the $|\cos(\theta^*)|$ and $\mathcal{A}$ variables is evaluated in data
and simulated multijet samples, where it amounts to 1.8\% and 2.2\%, respectively.
Significant correlations are  observed in data at large \mavg{} and high $\mathcal{A}$ values;
to reduce their impact on the background estimate
the $\mathcal{A}$--$|\cos(\theta^*)|$ plane is restricted to $0.0<\mathcal{A}<0.7$ and $0.0<|\cos(\theta^*)|<0.7$.
Two additional regions (B and E) are defined in the $\mathcal{A}$--$|\cos(\theta^*)|$  plane.
The validation region (VR), region E, is used to test the performance
of the data-driven method and assign an uncertainty to the background estimate.
The validation region is defined with the same selections as for the signal region, 
but with the asymmetry requirement changed from $\mathcal{A} < 0.05$ to $0.05 < \mathcal{A} < 0.15$.
The background contribution in the VR is estimated by $N_\mathrm{E} = N_\mathrm{B} \times N_\mathrm{F}/N_\mathrm{C}$.
In the inclusive selection the data-driven estimate also accounts for the contribution from the  $t\bar{t}$ production,
which  amounts to less than 1\% of the total background for $m_{\mathrm{avg}}<200$~\GeV, and is negligible above.

For the $b$-tagged selection, where the background is relatively small,
the signal contamination in region $\mathrm{A}$ can be significant
and potentially bias the result of the  background estimate.
The multijet background for this selection is thus estimated in two steps.
The shape of the \mavg{} distribution is first predicted in a region with a $b$-tag veto (zero-tag)
and then extrapolated to the $b$-tagged signal region.
The \mavg{} distribution in the zero-tag region is obtained with a data-driven estimate, analogously to the inclusive selection.
The zero-tag prediction  is then extrapolated to the $b$-tagged selection  by means of projection factors
computed bin by bin in \mavg, similarly to the approach described in Ref.~\cite{SUSY-2015-05}. 
The projection factors, for a given \mavg{} bin and region in the $\mathcal{A}$--$|\cos(\theta^*)|$  plane,
are defined as the ratio of the numbers of events with two $b$-tags and zero $b$-tags, $N_{\mathrm{two-}b\mathrm{-tags}}/N_{\mathrm{zero-}b\mathrm{-tags}}$, within that region.
The method assumes the projection factors to be constant across the $\mathcal{A}$--$|\cos(\theta^*)|$  plane.
They are evaluated in region F, where a negligible signal contamination is expected.
The contributions from multi-jet and $t\bar{t}$ production scale differently between the zero- and $b$-tagged selection.
Hence, simulated  samples are used to subtract the $t\bar{t}$ contribution in all control regions.
The $t\bar{t}$ estimate in the signal region is then obtained directly from the simulation,
considering all relevant modelling and experimental uncertainties.

\begin{table}[!ht]

  \caption{\label{tab:bkgdest:numbers}  Observed numbers of events and the predicted \ttbar contributions
    in each of the regions  used in the background estimate, for each $b$-tag multiplicity.
    The expected fractional signal contributions are shown for the mass windows
    corresponding to $m_{\tilde{t}}=125$~\GeV, 250~\GeV, 500~\GeV, and 800~\GeV{}.
    For the $m_{\tilde{t}}=125$~\GeV{} and 250~\GeV{} mass windows the fractions of \ttbar are also shown.
    The \ttbar  systematic uncertainties include both the detector-level uncertainties
    and the theoretical uncertainties, as described in Section~\ref{sec:systematics}.
  }
  \footnotesize
  
  \begin{center}
    \sisetup{group-minimum-digits=5}
    \scalebox{0.9}{  
      \begin{tabular}{
          c|
          S[table-format=7.0, table-number-alignment=center, round-mode=places, round-precision=0] |
          S[table-format=4.0, table-number-alignment=right, round-mode=places, round-precision=0]@{\quad$\pm$\quad}
          S[table-format=2.0, table-number-alignment=left, round-mode=places, round-precision=0]@{\quad$\pm$\quad}
          S[table-format=3.0, table-number-alignment=left, round-mode=figures, round-precision=2] |
          S[table-format=3.3,table-align-text-post=false] S[table-format=2.3,table-align-text-post=false] |
          S[table-format=3.3,table-align-text-post=false]  S[table-format=2.3,table-align-text-post=false]
          | S[table-format=2.2,table-align-text-post=false]  | S[table-format=1.1,table-align-text-post=false] }
    \toprule
    \multicolumn{5}{c|}{Target mass} & \multicolumn{2}{c|}{$125~\gev$} & \multicolumn{2}{c|}{$250~\gev$} & \multicolumn{1}{c|}{$500~\gev$} & \multicolumn{1}{c}{$800~\gev$} \\
    Region & $N_{\mathrm{Data}}$ & \multicolumn{3}{c|}{$N_{\ttbar}$ ($\pm$ stat. $\pm$ syst.)} & \multicolumn{2}{c|}{$[120, 135]~\gev$} & \multicolumn{2}{c|}{$[230, 260]~\gev$} & \multicolumn{1}{c|}{$[455, 515]~\gev$} & \multicolumn{1}{c}{$[720, 820]~\gev$} \\
    & & \multicolumn{3}{c|}{} & $\frac{N_{\mathrm{Sig}}}{N_{\mathrm{Data}}}$ & $\frac{N_{\ttbar}}{N_{\mathrm{Data}}}$ & $\frac{N_{\mathrm{Sig}}}{N_{\mathrm{Data}}}$ & $\frac{N_{\ttbar}}{N_{\mathrm{Data}}}$ & $\frac{N_{\mathrm{Sig}}}{N_{\mathrm{Data}}}$ & $\frac{N_{\mathrm{Sig}}}{N_{\mathrm{Data}}}$\\

    \midrule
    \multicolumn{11}{c}{Inclusive selection}\\
    \midrule                                           
    A     & 256937     & 5044  & 76  & 1092   & 7.2\%  & 5.8\%  & 5.6\% & 0.28\% & 3.1\% &  1.7 \%  \\
    B     & 508589     & 8900  & 100 & 1410   & 1.95\%  & 4.7\% & 1.3\% & 0.24\% & 0.6\% &  0.4 \%  \\
    C     & 1154721    & 13080 & 120 & 1950   & 0.17\%  & 2.3\% & 0.16\% & 0.43\% & 0.07\% &  0.07 \%  \\
    D (SR)& 154750     & 3826  & 66  & 812    & 14.0\%  & 7.0\% & 10.5\% & 0.31\% & 6.3\% &  3.5 \%  \\
    E (VR)& 307268     & 6578  & 87  & 995    & 3.86\%  & 6.0\% & 2.2\% & 0.33\% & 1.4\% &  0.8 \%  \\
    F     & 694492     & 9920  & 110 & 1900   & 0.29\%  & 3.5\% & 0.3\% & 0.57\% & 0.2\% &  0.13 \%  \\
    \midrule
    \multicolumn{11}{c}{Zero $b$-tags selection}\\
    \midrule
    A     & 184432 & 580  & 27 & 85   & 0.56\% & 0.53\% & 0.44\% & 0.11\% & 0.50\% &  0.46 \% \\
    B     & 366003 & 1165 & 38 & 213  & 0.14\% & 0.57\% & 0.18\% & 0.10\% & 0.12\% &  0.19 \% \\
    C     & 834944 & 2352 & 53 & 399  & 0.07\% & 0.66\% & 0.03\% & 0.13\% & 0.02\% &  0.04 \% \\
    D     & 110071 & 506  & 26 & 94   & 1.18\% & 0.72\% & 1.65\% & 0.16\% & 1.48\% &  1.3 \% \\
    E     & 219366 & 831  & 32 & 183  & 0.45\% & 0.67\% & 0.24\% & 0.11\% & 0.10\% &  0.4 \% \\
    F     & 498751 & 1743 & 46 & 291  & 0.07\% & 0.83\% & 0.08\% & 0.20\% & 0.08\% &  0.04 \% \\
    \midrule 
    \multicolumn{11}{c}{$b$-tagged selection}\\
    \midrule
    A      & 8484    & 2375 & 53 & 902 & 82\%  & 64\%  & 112\% & 0.94\%   & 43\% & 18.2 \% \\  
    B      & 16113   & 3614 & 64 & 867 & 23\%  & 53\%  & 23\% & 1.4\%     & 11\% & 4.5 \% \\
    C      & 32759   & 3681 & 63 & 840 & 1.2\% & 31\%  & 1.3\% & 2.2\%    & 0.38\% & 0.1 \% \\
    D (SR) & 5603    & 1707 & 44 & 499 & 135\% & 64\%  & 181\% & 0.54\%   & 70\% & 24.6 \% \\
    E (VR) & 10531   & 2678 & 55 & 499 & 38\%  & 58\%  & 35\% & 0.9\%     & 20\% & 9.2 \% \\
    F      & 20856   & 2904 & 56 & 721 & 2.3\% & 37\%  & 3.1\% & 2.7\%    & 1.4\% & 0.7 \% \\
    \bottomrule
      \end{tabular}
      }
  \end{center}
\end{table}

\begin{figure}
  \centering
  \includegraphics[width=0.49\textwidth]{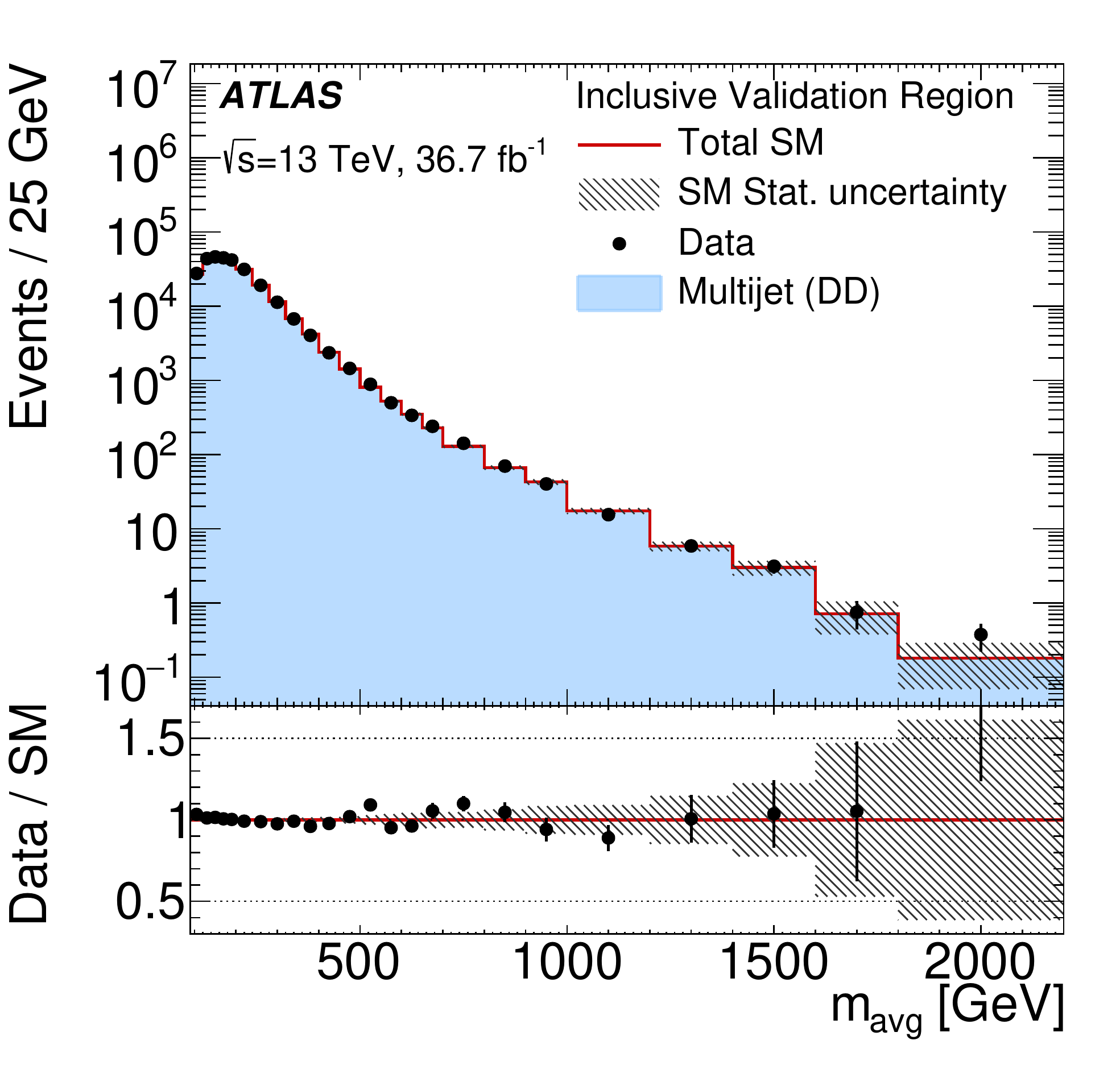}
  \includegraphics[width=0.49\textwidth]{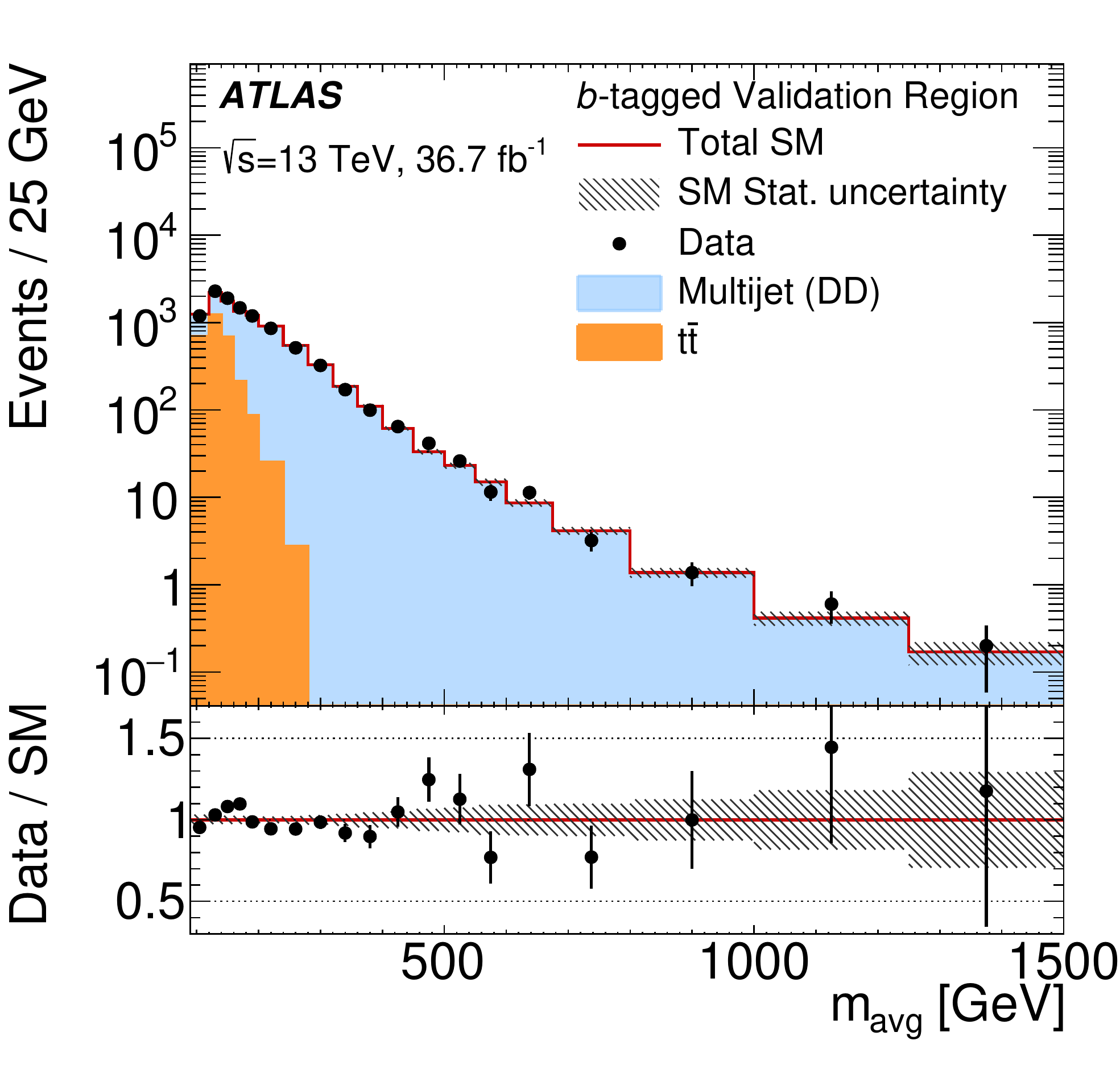}
  \caption{\label{fig:validation_regions}  The \mavg{} spectrum in the   inclusive (left) and $b$-tagged (right) validation regions.
    The data (black points) are compared to the total background prediction (red line) estimated with the data-driven method.
    The fraction of background coming from top-pair production is shown in orange.
    The statistical uncertainties of the prediction are shown by the grey hatched band.}

\end{figure}

The observed number of events in each of the regions  used in the background estimate before  the mass window requirements are applied,
together with the expected signal contamination in few representative mass windows,
are shown for both the inclusive and $b$-tagged selections in Table~\ref{tab:bkgdest:numbers}.
The \mavg{} distribution in the validation region for the inclusive and $b$-tagged signal regions is shown in Figure~\ref{fig:validation_regions}.
Within the statistical uncertainties the method  reproduces both the normalisation and the shape of \mavg{} in the VRs.
The level of agreement observed in the VRs is used to derive a systematic uncertainty in the background estimate in the SR.
In each \mavg{} mass window the difference between the observed data and the estimation  in the VR (non-closure) is computed.
The larger of the observed non-closure in the VR 
and the  statistical uncertainty of the data-driven method is assigned as an uncertainty in the background estimates.
To reduce the effect of statistical fluctuations in the non-closure
and avoid  quoting an unphysically small value of the systematic uncertainty for the mass windows where it changes sign,
this  uncertainty is further smoothed as a function of \mavg.
The Nadaraya--Watson~\cite{nadaraya1964estimating,watson1964smooth} kernel regression estimate is used for the smoothing,
with a bandwidth of 500~\GeV{} (meaning that the quartiles of the kernels are placed at  $\pm125$~\GeV).
The uncertainties assigned to the background estimate in the inclusive and $b$-tagged signal regions are summarised in Figure~\ref{fig:closure}.

\begin{figure}
  \centering
  \includegraphics[width=0.49\textwidth]{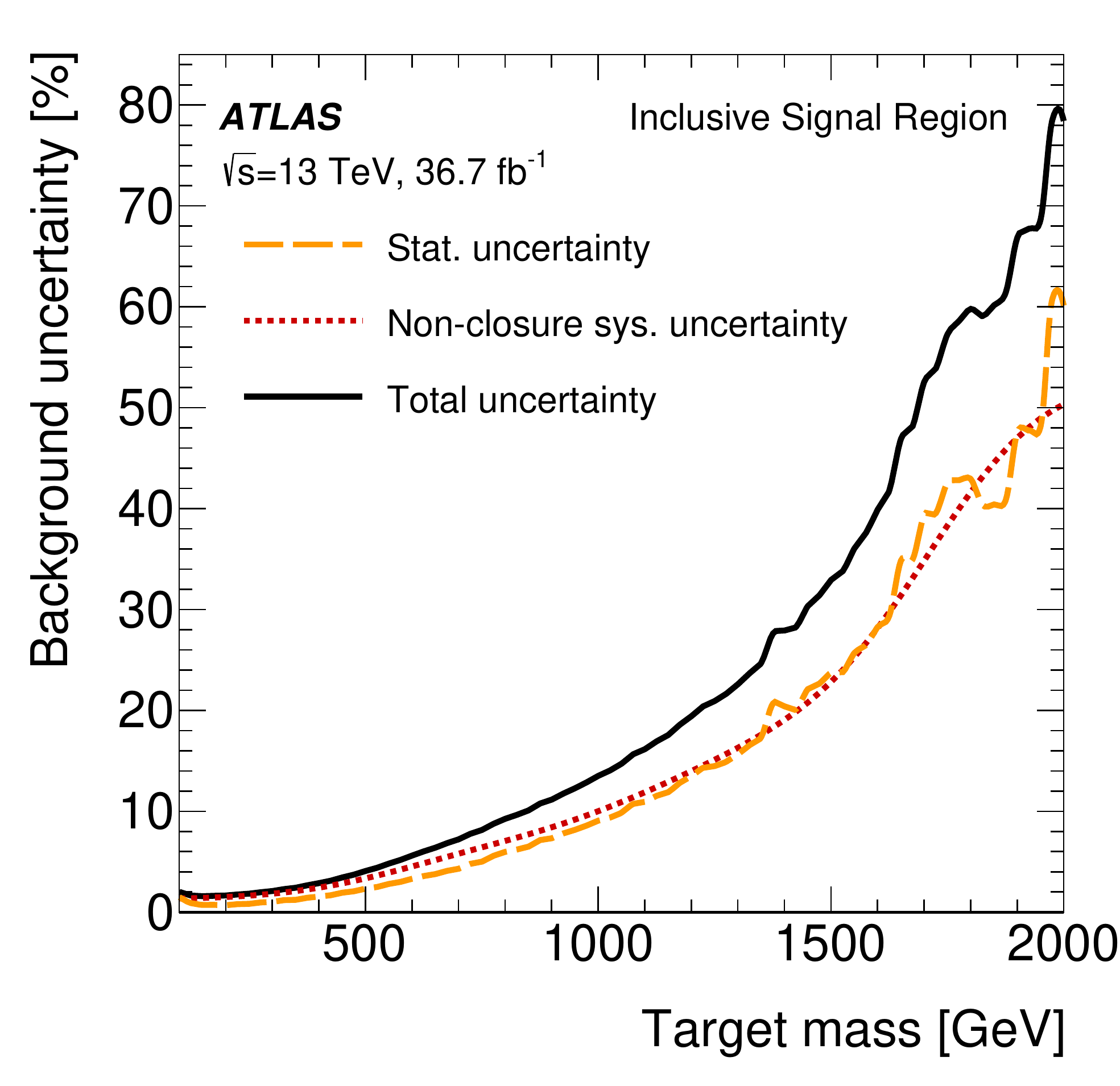}
  \includegraphics[width=0.49\textwidth]{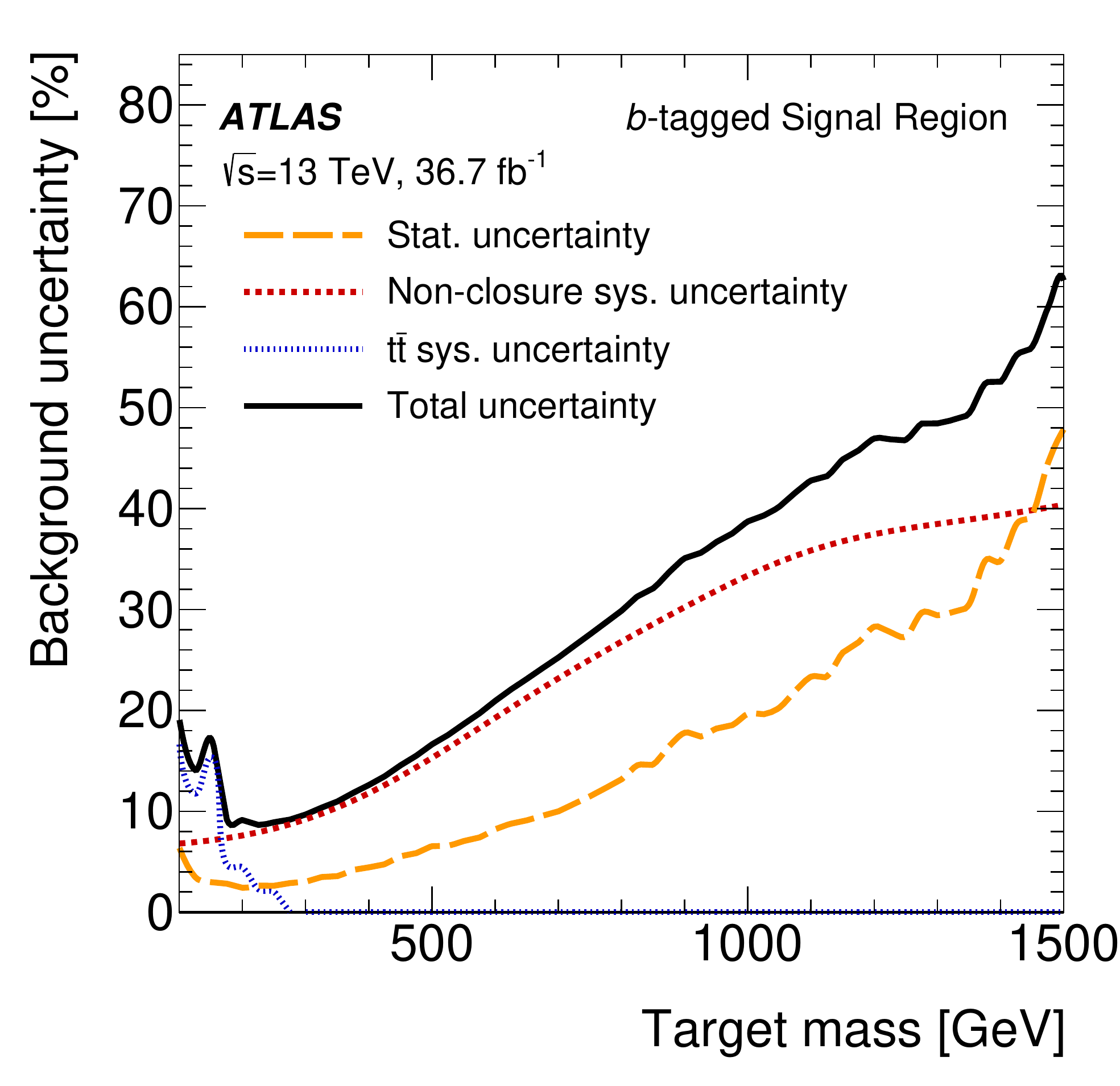}\\
  \caption{\label{fig:closure}  The  uncertainty in the data-driven background estimate in 
    the inclusive (left) and $b$-tagged (right) signal regions,
    computed in the \mavg{} mass windows defined for different target masses.
    The uncertainty arising from the non-closure in the validation region is shown with a red short-dashed line
    and compared with the statistical uncertainty of the data-driven prediction shown as an orange dashed line.
    The additional uncertainty in the MC estimate of the top background in the $b$-tagged signal region is shown as a dotted blue line.
    The total uncertainty, obtained by adding in quadrature the different components, is indicated by a black solid line.
  }
\end{figure}

\section{Systematic uncertainties}\label{sec:systematics}
While the multijet background uncertainties pertain primarily to
the estimation method itself, the top background and the signals  are also affected by uncertainties related to the
description of detector effects and to the physics modelling of the MC simulation.

The dominant detector-related systematic
effects are due to the uncertainties in the jet energy scale \cite{ATL-PHYS-PUB-2015-015}
and resolution \cite{ATLAS-CONF-2015-037} and from the  $b$-tagging efficiency and mistag rate~\cite{ATL-PHYS-PUB-2016-012}.

Since MC simulation is used to determine the contribution from top events in the $b$-tagged signal region,
systematic uncertainties related to the choice of MC generator for the process need to be estimated.
These are evaluated by comparing the nominal samples to additional samples with systematic variations.
A modelling uncertainty  is derived by comparing the predictions 
of the nominal sample with a sample produced with \textsc{Powheg} interfaced with \textsc{Herwig++} 2.7.1,
or with \textsc{MG5}\_a\textsc{MC@NLO} interfaced with \textsc{Herwig++}.
In addition, the difference in the prediction obtained by modifying the parton-shower intensity
and the $h_{\mathrm{damp}}$ parameter in the nominal  sample is taken as an uncertainty.
The  $t\bar{t}$ systematic uncertainty on the total background is as large as 20\% for \mavg{} below 200~\GeV,
becoming negligible above 200~\GeV.

The total detector-related uncertainties in the signal are about 10\% for the inclusive SR
and  about 15\% for the two-$b$-tagged SR.
For top squark production the nominal signal cross-sections and their uncertainties are taken from an
envelope of cross-section predictions derived using different PDF sets and different factorisation and
renormalisation scales, as described in Ref.~\cite{Borschensky:2014cia}. The theoretical uncertainties in the acceptance of the
signal simulation include variations of the renormalisation and factorisation scales, the CKKW-L merging
scales, and the value of the strong coupling constant in \textsc{MG5}\_a\textsc{MC@NLO} as well as parton shower uncertainties in \textsc{Pythia} 8
evaluated from variations of the A14 parameter set.
After normalising the samples using the same cross-section,
the difference between the yields from the nominal and varied samples
in the mass window, which is typically below 1\%, is considered as an uncertainty.

\section{Results and interpretation}
The \mavg{} distributions in the inclusive and $b$-tagged signal regions are shown in Figure~\ref{fig:signal_regions}. 
Agreement is observed between  data and the expected background.
The expected numbers of background and signal events in the SR and their uncertainties
are reported  for the  mass windows defined for top squark and coloron signals 
in Tables~\ref{tab:totalyields_inc} and \ref{tab:totalyields_inc_col}, respectively,
together with the observed events in data.
Table~\ref{tab:totalyields_2b} presents the numbers in the top squark mass windows of the two-$b$-tagged signal region.

To estimate the compatibility of the data with a generic resonance mass hypothesis,
the \mavg{} distribution is scanned in 12.5~\GeV{} steps.
The \mavg{} window for an arbitrary mass is obtained from a linear fit
to the lower and upper edges of the windows obtained for the simulated signal masses.
For each mass a  background $p_0$-value is computed  for the inclusive and $b$-tagged signal regions.
The largest deviation is found in the $b$-tagged signal region for a mass of 463~\GeV,
corresponding to a local $p_0$-value of 0.05.

The expected $p_0$-values in each mass window are also evaluated for potential signals.
At least three-standard-deviation $(3\sigma)$ signal sensitivity is expected for top squark masses up to 350~\GeV{} 
with the inclusive signal region and 450~\GeV{} with the $b$-tagged signal region.
For colorons a greater than $3\sigma$ sensitivity is expected for masses up to 1400~\GeV{}.

\begin{table}\centering
  \caption{\label{tab:totalyields_inc}
    Observed numbers of events in the data, $N_{\mathrm{Data}}$,
    the estimated numbers of background events, $N_{\mathrm{Bkg}}$,
    and the expected numbers of top squark signal events, $N_{\mathrm{Sig}}$,
    in the top squark mass windows of the inclusive signal region.
    Separate statistical and systematic uncertainties are given.}
  \sisetup{round-mode=figures, round-precision=2, retain-explicit-plus=true, group-four-digits = true,group-minimum-digits=5}

  \footnotesize
  \begin{tabular}{r|c|S[table-format=5.0,round-mode=places, round-precision=0]|
      S[table-format=6.0, round-mode=places, round-precision=3]@{$\pm$}
      S[table-format=3.0,round-mode=places, round-precision=2]@{$\pm$}
      S[table-format=3.0,round-mode=places, round-precision=2]|
      S[table-format=5.1,round-mode=places, round-precision=1]@{$\pm$}
      S[table-format=2.1,round-mode=places, round-precision=1]@{$\pm$}
      S[table-format=3.1,round-mode=places, round-precision=1]
    }
    \toprule
    $m_{\tilde{t}}$ [GeV] & Window [\GeV] & $N_{\mathrm{Data}}$ & \multicolumn{3}{c|}{$N_{\mathrm{Bkg}}$ ($\pm$ stat. $\pm$ syst.)} & \multicolumn{3}{c}{$N_{\mathrm{Sig}}$ ($\pm$ stat. $\pm$ syst.)} \\                   
    \midrule       
 100    &  [100, 110]  &  5899   &   5910    &  90   &   70    &  519    &   23     &   68 \\ 
 125    &  [120, 135]  &  13497   &   13450    &  120   &   180   &  1890    &   50     &   190  \\ 
 150    &  [140, 160]  &  18609   &   18390    &  130   &   250   &  2540    &   50     &   130  \\ 
 175    &  [165, 185]  &  17742   &   17800    &  130   &   250   &  2280    &   50     &   210  \\ 
 200    &  [185, 210]  &  19844   &   19660    &  140   &   290   &  2250    &   50     &   170  \\ 
 225    &  [210, 235]  &  14898   &   15180    &  120   &   230   &  1620    &   40     &   100  \\ 
 250    &  [230, 260]  &  13689   &   13750    &  110   &   220   &  1440    &   80     &   140  \\ 
 275    &  [255, 285]  &  9808   &   9860    &  100   &   170   &  1010    &   70     &   80  \\ 
 300    &  [275, 310]  &  8514   &   8790    &  90   &   160   &  789    &   52     &   31 \\ 
 325    &  [300, 335]  &  6180   &   6330    &  80   &   120   &  600    &   50     &   50 \\ 
 350    &  [320, 365]  &  5802   &   5900    &  70   &   120   &  509    &   39     &   19 \\ 
 375    &  [345, 390]  &  4113   &   4250    &  60   &   90   &  324    &   25     &   31 \\ 
 400    &  [365, 415]  &  3531   &   3590    &  60   &   90   &  274    &   14     &   18 \\ 
 425    &  [385, 440]  &  3108   &   3010    &  50   &   80   &  198    &   23     &   10 \\ 
 450    &  [410, 465]  &  2281   &   2230    &  40   &   60   &  154    &   17     &   27 \\ 
 475    &  [430, 490]  &  1906   &   1920    &  40   &   60   &  116    &   12     &   8 \\ 
 500    &  [455, 515]  &  1495   &   1513    &  35   &   49   &  94    &   10     &   8 \\ 
 525    &  [475, 540]  &  1318   &   1327    &  33   &   46   &  71    &   7     &   4 \\ 
 550    &  [500, 565]  &  1050   &   1048    &  29   &   39   &  48.5    &   5.4     &   2.2 \\ 
 575    &  [520, 590]  &  924   &   912    &  27   &   36   &  44    &   4     &   4 \\ 
 600    &  [545, 620]  &  745   &   744    &  25   &   31   &  36.9    &   1.6     &   2.3 \\ 
 625    &  [565, 645]  &  645   &   626    &  22   &   28   &  30.3    &   2.8     &   3.4 \\ 
 650    &  [585, 670]  &  536   &   554    &  21   &   26   &  23.3    &   2.1     &   1.9 \\ 
 675    &  [610, 695]  &  438   &   473    &  19   &   24   &  20.3    &   1.6     &   0.9 \\ 
 700    &  [630, 720]  &  404   &   422    &  18   &   22   &  15.4    &   1.2     &   0.9 \\ 
 725    &  [655, 745]  &  341   &   335    &  16   &   18   &  13.6    &   1.0     &   0.9 \\ 
 750    &  [675, 770]  &  306   &   310    &  16   &   18   &  12.4    &   0.9     &   0.9 \\ 
 775    &  [700, 795]  &  265   &   243    &  14   &   14   &  9.7    &   0.7     &   0.7 \\ 
 800    &  [720, 820]  &  238   &   205    &  12   &   13   &  8.5    &   0.6     &   0.6 \\ 
    \bottomrule
  \end{tabular}
\end{table}

\begin{table}\centering
  \caption{\label{tab:totalyields_inc_col}
    Observed numbers of events in the data, $N_{\mathrm{Data}}$,
    the estimated numbers of background events, $N_{\mathrm{Bkg}}$,
    and the expected numbers of coloron signal events, $N_{\mathrm{Sig}}$,
    in the coloron mass windows of the inclusive signal region.
    Separate statistical and systematic uncertainties are given.}
  \sisetup{round-mode=figures, round-precision=2, retain-explicit-plus=true, group-digits = true,group-minimum-digits=5}
  \footnotesize
  \begin{tabular}{r|c|S[table-format=5.0,round-mode=places, round-precision=0]|
      S[table-format=5.1,round-mode=places, round-precision=1]@{$\pm$}
      S[table-format=2.1,round-mode=places, round-precision=1]@{$\pm$}
      S[table-format=2.1,round-mode=places, round-precision=1]|
      S[table-format=5.1,round-mode=places, round-precision=1]@{$\pm$}
      S[table-format=4.1,round-mode=places, round-precision=1]@{$\pm$}
      S[table-format=4.1,round-mode=places, round-precision=1]
    }
    \toprule
    $m_{\rho}$ [GeV] & Window [\GeV] & $N_{\mathrm{Data}}$ & \multicolumn{3}{c|}{$N_{\mathrm{Bkg}}$ ($\pm$ stat. $\pm$ syst.)} & \multicolumn{3}{c}{$N_{\mathrm{Sig}}$ ($\pm$ stat. $\pm$ syst.)} \\

   \midrule
    500    &  [455, 515]  &  1495   &   1513    &  35   &   15   &  23000    &   1900     &   1200 \\ 
    625    &  [565, 645]  &  645    &   626  &  22  &  35  &  7050  &  370  &  350  \\
    750    &  [675, 770]  &  306    &   310    &  15   &   30   &  2510    &   170     &   120 \\ 
    875    &  [790, 900]  &  166    &   144    &  10   &   16   &  1020    &   56     &   23 \\ 
    1000   &  [900, 1025] &  79     &   96    &  9   &   8   &  416    &   25     &   17 \\ 
    1125   &  [1010, 1155]  &  46   &   58    &  7   &   5   &  154    &   8     &   5 \\ 
    1250   &  [1120, 1280]  &  27   &   36    &  5   &   3   &  73    &   4     &   4 \\ 
    1375   &  [1235, 1410]  &  9    &   17    &  3  &  3  &  51.0  &  2.0  &  1.2  \\
    1500   &  [1345, 1535]  &  13   &   14    &  3   &   1.6   &  12.9    &   0.8     &   0.4 \\ 
    1625   &  [1455, 1665]  &  7    &   8.70  &  2.56   &   0.6   &  12.9    &   0.8     &   0.4 \\ 
    1750   &  [1565, 1790]  &  6    &   4.79  &  2.04  &  2.55  &  2.80  &  0.12  &  0.13  \\
    1875   &  [1680, 1920]  &  4    &   5.27  &  2.15  &  3.51  &  1.33  &  0.07  &  0.07  \\
    2000   &  [1790, $\infty$]  &  2   &   2.07    &  1.24   &   0.4   &  0.64    &   0.06     &   0.06 \\ 
    \bottomrule
  \end{tabular}
\end{table}

\begin{table}\centering
  \caption{\label{tab:totalyields_2b}
    Observed numbers of events in the data, $N_{\mathrm{Data}}$,
    the estimated numbers of background events, $N_{\mathrm{Bkg}}$,
    and the expected numbers of top squark signal events, $N_{\mathrm{Sig}}$,
    in the top squark mass windows of the $b$-tagged signal region.
    Separate statistical and systematic uncertainties are given.}
  \footnotesize
  \sisetup{round-mode=figures, round-precision=2, retain-explicit-plus=true, group-digits = true,group-minimum-digits=5}
  \begin{tabular}{r|c|S[table-format=4.0,round-mode=places, round-precision=0]|
      S[table-format=4.1,round-mode=places, round-precision=1]@{$\pm$}
      S[table-format=2.1,round-mode=places, round-precision=1]@{$\pm$}
      S[table-format=2.1,round-mode=places, round-precision=1]|
      S[table-format=4.1,round-mode=places, round-precision=1]@{$\pm$}
      S[table-format=2.1,round-mode=places, round-precision=1]@{$\pm$}
      S[table-format=2.1,round-mode=places, round-precision=1]
    }
    \toprule
    $m_{\tilde{t}}$ [\GeV] & Window [\GeV] & $N_{\mathrm{Data}}$ & \multicolumn{3}{c|}{$N_{\mathrm{Bkg}}$ ($\pm$ stat. $\pm$ syst.)} & \multicolumn{3}{c}{$N_{\mathrm{Sig}}$ ($\pm$ stat. $\pm$ syst.)} \\
    \midrule
 100    &  [100, 110]  &  256   &   285    &  18   &   51   &  308    &   18     &   52 \\ 
 125    &  [120, 135]  &  803   &   798    &  28   &   107   &  1090    &   40     &   140 \\ 
 150    &  [140, 160]  &  809   &   789    &  23   &   132   &  1510    &   40     &   130 \\ 
 175    &  [165, 185]  &  544   &   555    &  16   &   47   &  1300    &   40     &   140 \\ 
 200    &  [185, 210]  &  592   &   554    &  13   &   47   &  1220    &   40     &   110 \\ 
 225    &  [210, 235]  &  414   &   436    &  11   &   35   &  893    &   28     &   90 \\ 
 250    &  [230, 260]  &  416   &   385    &  10   &   32   &  750    &   60     &   120 \\ 
 275    &  [255, 285]  &  302   &   283    &  8   &   24   &  480    &   50     &   60 \\ 
 300    &  [275, 310]  &  242   &   250    &  8   &   23   &  390    &   40     &   50 \\ 
 325    &  [300, 335]  &  181   &   179    &  6   &   17   &  273    &   33     &   34 \\ 
 350    &  [320, 365]  &  169   &   161    &  6   &   16   &  225    &   25     &   20 \\ 
 375    &  [345, 390]  &  110   &   111    &  5   &   12   &  147    &   16     &   22 \\ 
 400    &  [365, 415]  &  80   &   96    &  4   &   11   &  114    &   9     &   12 \\ 
 425    &  [385, 440]  &  85   &   79    &  4   &   10   &  76    &   14     &   11 \\ 
 450    &  [410, 465]  &  71   &   54.2    &  3.0   &   7.1   &  48    &   9     &   10 \\ 
 475    &  [430, 490]  &  67   &   46.8    &  2.7   &   6.5   &  40    &   7     &   5 \\ 
 500    &  [455, 515]  &  38   &   35.8    &  2.3   &   5.3   &  26    &   5     &   5 \\ 
 525    &  [475, 540]  &  31   &   35.1    &  2.3   &   5.5   &  21.7    &   3.9     &   2.8 \\ 
 550    &  [500, 565]  &  20   &   30.2    &  2.1   &   5.0   &  12.4    &   2.5     &   2.3 \\ 
 575    &  [520, 590]  &  14   &   26.3    &  2.0   &   4.6   &  17.5    &   2.7     &   3.5 \\ 
 600    &  [545, 620]  &  14   &   19.5    &  1.6   &   3.5   &  11.4    &   0.9     &   1.5 \\ 
 625    &  [565, 645]  &  15   &   15.8    &  1.4   &   3.0   &  9.3    &   1.5     &   1.4 \\ 
 650    &  [585, 670]  &  14   &   14.6    &  1.3   &   2.9   &  6.9    &   1.2     &   1.1 \\ 
 675    &  [610, 695]  &  13   &   13.6    &  1.3   &   2.8   &  5.5    &   0.8     &   0.6 \\ 
 700    &  [630, 720]  &  6   &   12.1    &  1.2   &   2.6   &  4.3    &   0.6     &   0.5 \\ 
 725    &  [655, 745]  &  5   &   9.9    &  1.1   &   2.2   &  4.4    &   0.6     &   0.8 \\ 
 750    &  [675, 770]  &  4   &   8.4    &  0.1   &   1.9   &  3.4    &   0.5     &   0.5 \\ 
 775    &  [700, 795]  &  8   &   6.9    &  0.9   &   1.6   &  2.36    &   0.34     &   0.53 \\ 
 800    &  [720, 820]  &  7   &   5.3    &  0.7   &   1.3   &  1.72    &   0.26     &   0.23 \\ 

    \bottomrule
  \end{tabular}
\end{table}

\begin{figure}
  \centering
\includegraphics[width=0.49\textwidth]{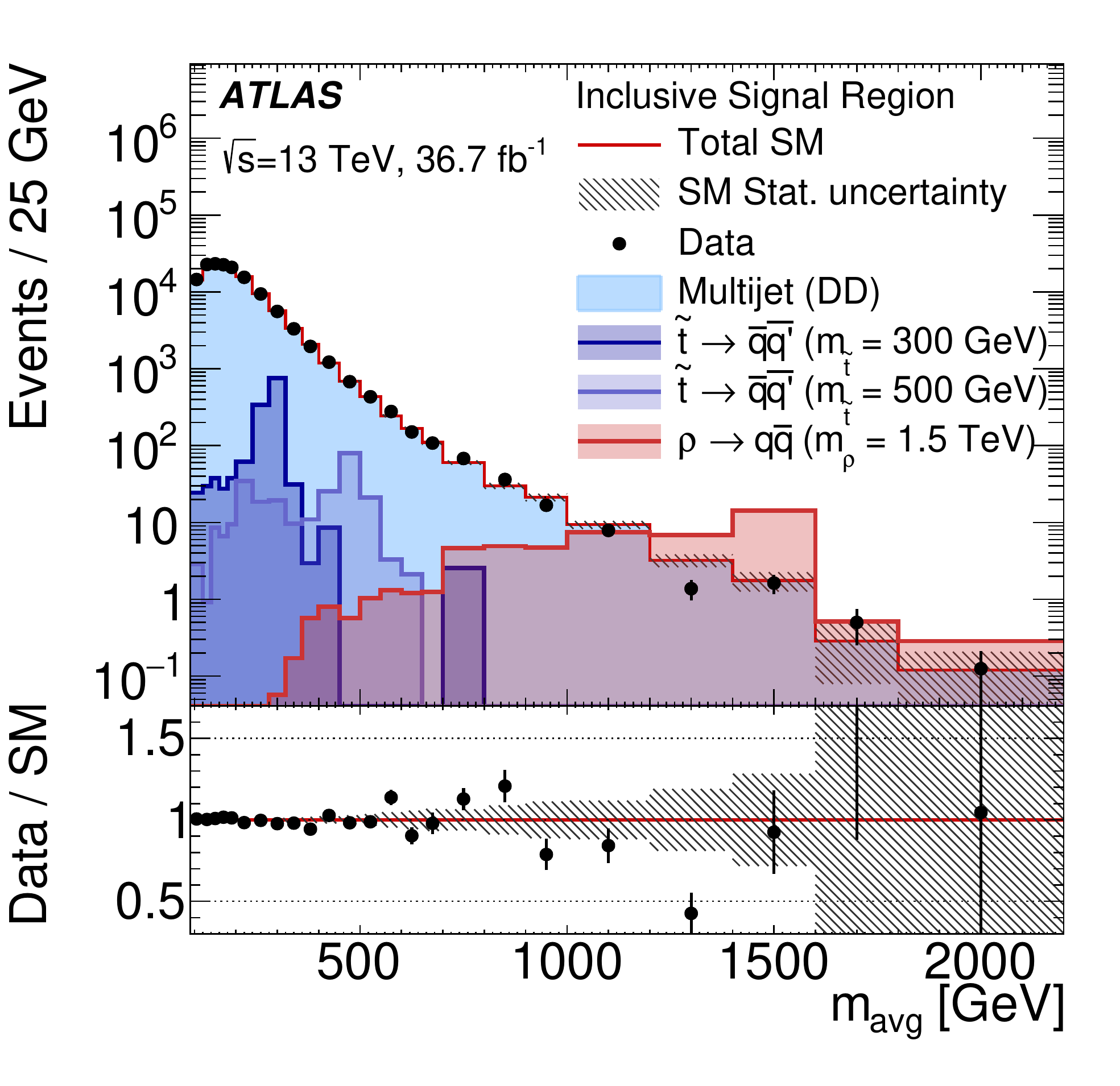}
  \includegraphics[width=0.49\textwidth]{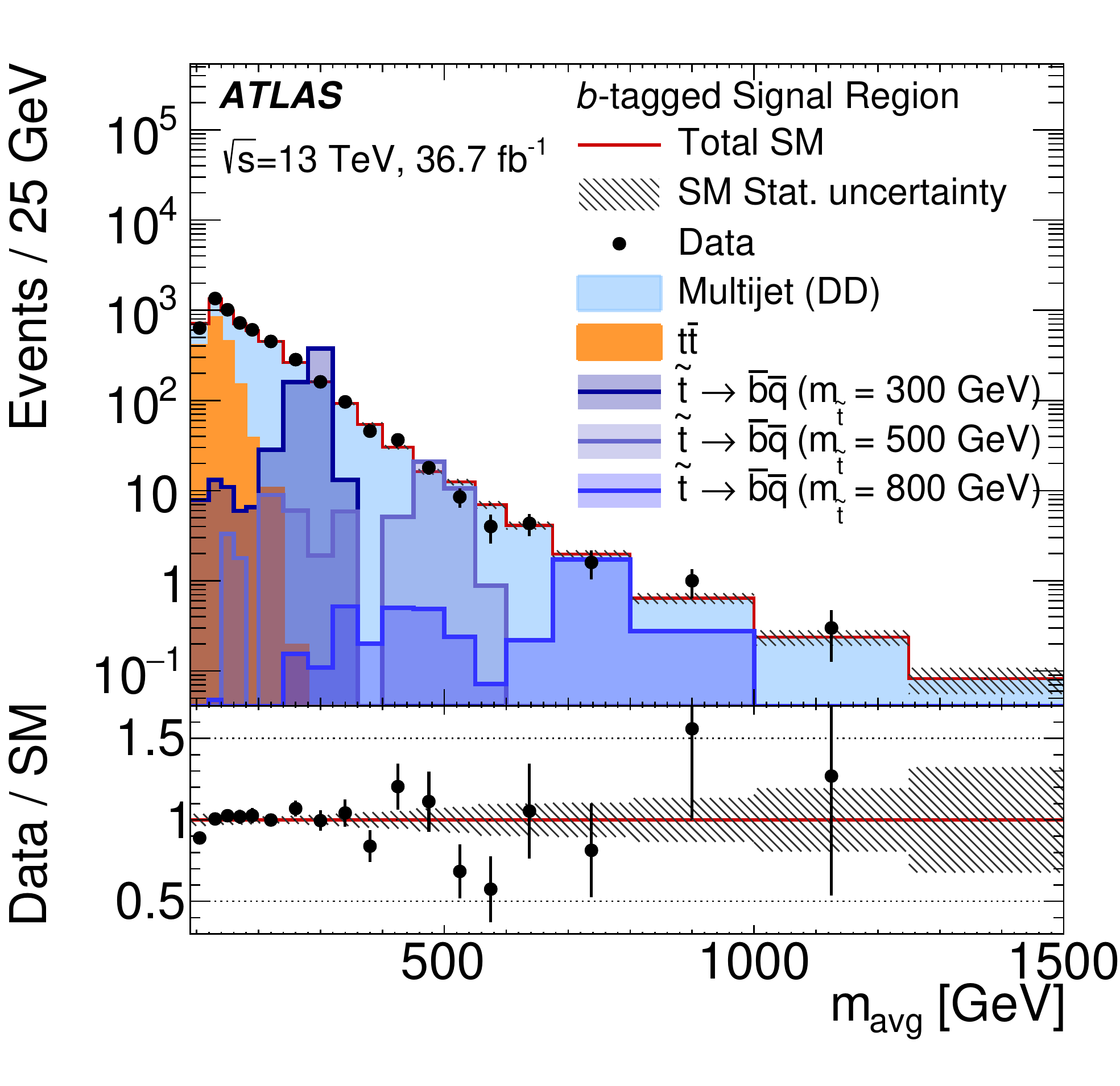}
  \caption{\label{fig:signal_regions}  The \mavg{} spectrum in the  inclusive (left) and $b$-tagged (right) signal regions.
    The data (black points) are compared to the total background prediction (red line) estimated with the data-driven method.
    The fraction of background coming from top-pair production is shown in orange. 
    The statistical uncertainties of the prediction are shown by the grey hatched band.
    Signals of different masses are overlaid in different colours.}
\end{figure}

In the absence of a statistically significant excess in data, exclusion limits are derived for the investigated signal models.
The inclusive signal region is used to set a limit on top squark, sgluon and coloron production with decays into a pair of jets,
while the $b$-tagged signal region is used to interpret the search for top squark pair production with decays into a $b$- and a light-quark jet.
A profile likelihood ratio combining Poisson probabilities for signal and background is computed
to determine the 95\% CL interval for compatibility of the data with the signal-plus-background hypothesis ($\mathrm{CL}_\mathrm{s+b}$) \cite{Cowan:2010js}.
A similar calculation is performed for the background-only hypothesis ($\mathrm{CL}_\mathrm{b}$).
From the ratio of these two quantities, the confidence level for the presence of a signal ($\mathrm{CL}_\mathrm{s}$) is determined \cite{Read:2002hq}.
Systematic uncertainties are treated as nuisance parameters and are assumed to follow Gaussian distributions.
The results are evaluated using pseudo-experiments.
This procedure is implemented using a software framework for statistical data analysis, HistFitter \cite{Baak:2014wma}.
The observed and expected 95\% CL upper limits on the allowed cross-sections are shown in Figure~\ref{fig:limits}.
For top squark  decays into two quarks, the expected and observed mass range exclusions
are between 100~\GeV{} to 430~\GeV{} and between 100~\GeV{} and 410~\GeV, respectively.
This exclusion does also  apply to the pair-production of other squarks,  decaying, for example, to a $d$- and a $u$-quark.
If the top squark decay is into a $b$-quark and a light-quark,
masses between 100~\GeV{} and 530~\GeV{} are expected to be excluded,
with the observed exclusion ranging from 100~\GeV{} to 470~\GeV{} and from 480~\GeV{} to 610~\GeV.
Below top squark masses of about 200~\GeV{} the 
signal acceptance rapidly drops due to the trigger and jet requirements,
and the analysis sensitivity does not surpass the 8~\TeV{} result,
which was specifically optimised for low-mass signals.
Pair-produced scalar gluons with decays into two gluons are excluded up to a mass of 800~\GeV{}.
Pair-produced colorons coupling only to light quarks are excluded  up to a mass of 1500~\GeV.

\begin{figure}
  \centering
  \subfloat[]{\includegraphics[width=0.49\textwidth]{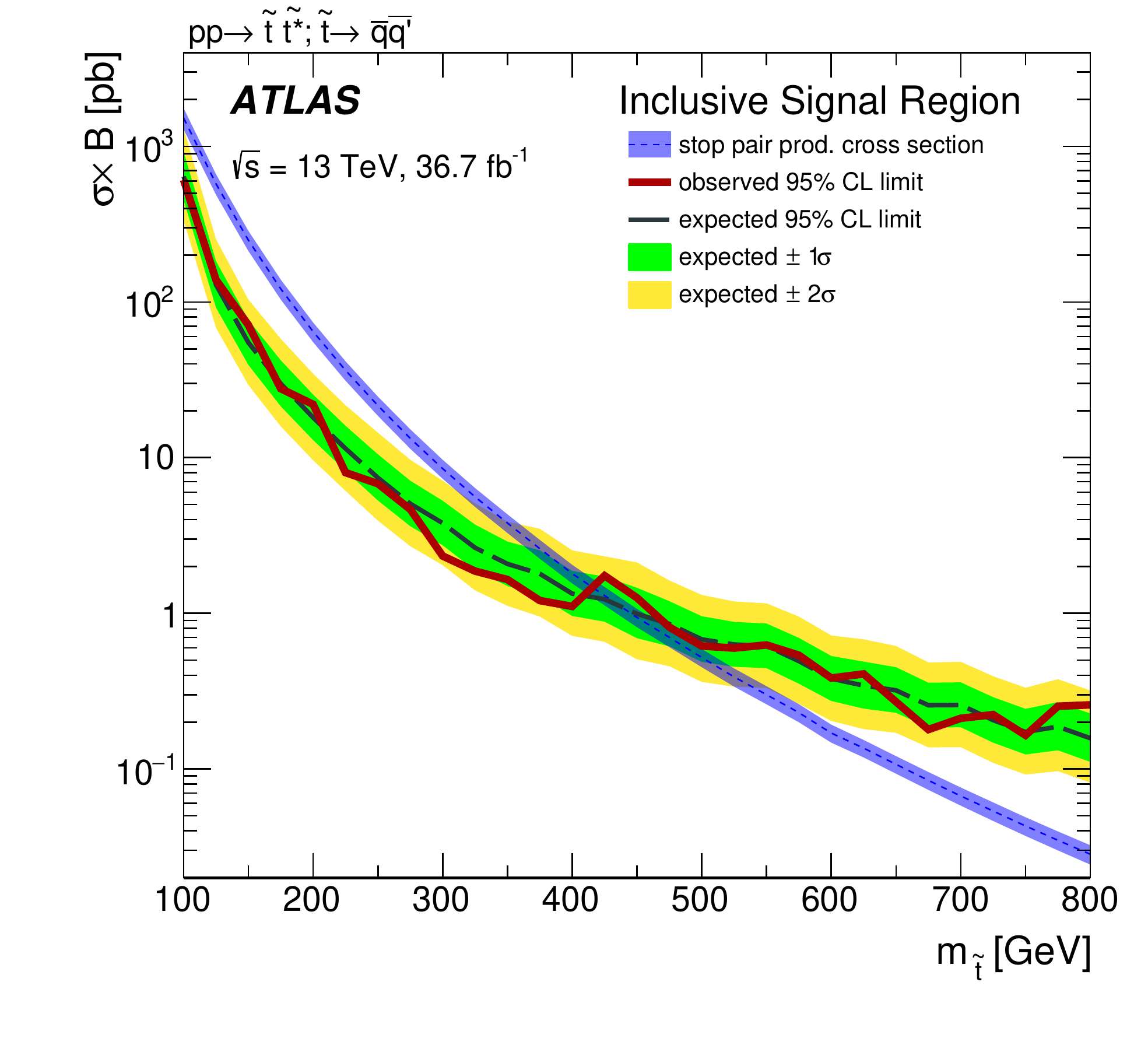}\label{subfig:limits_inc}}
  \subfloat[]{\includegraphics[width=0.49\textwidth]{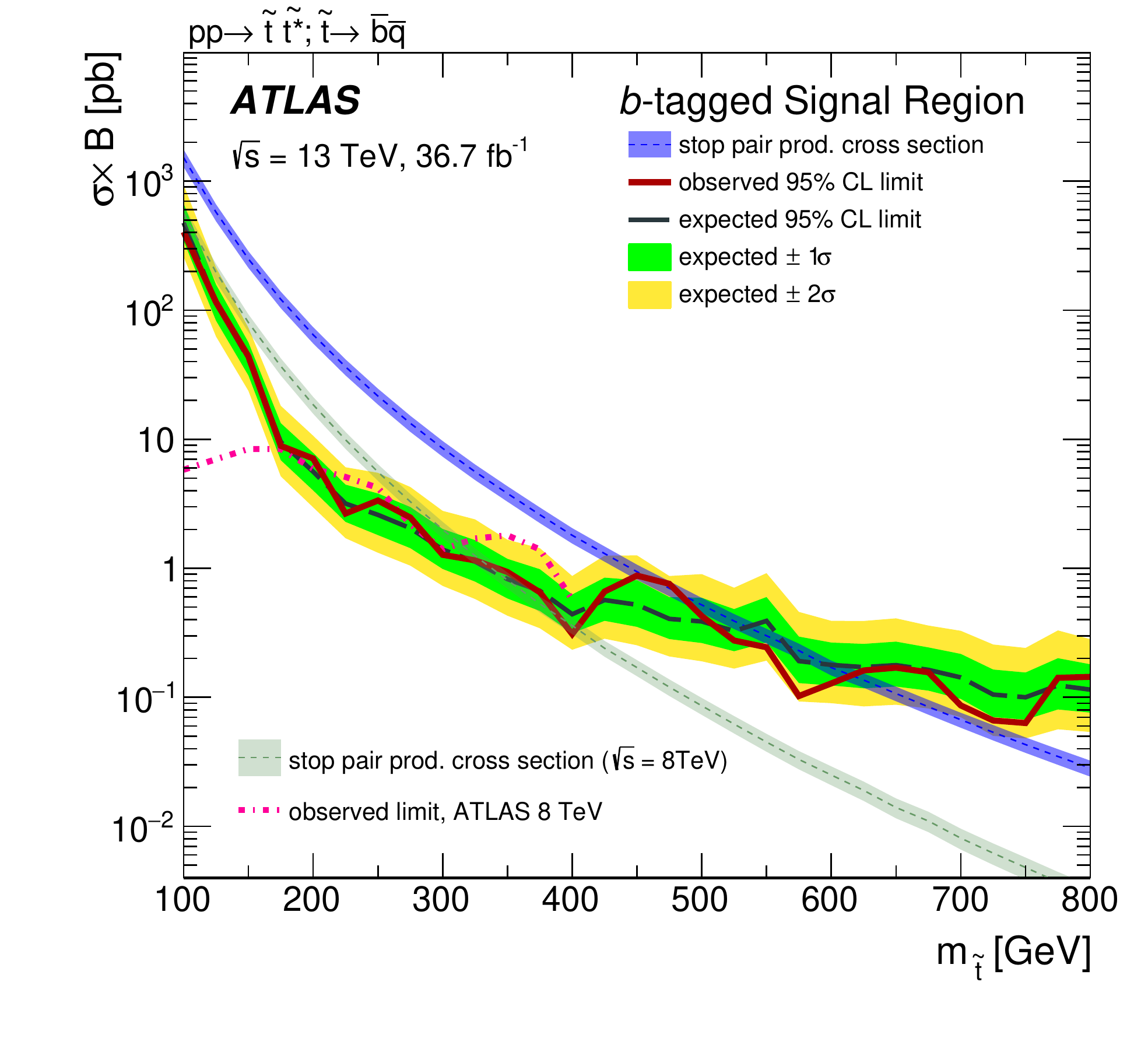}\label{subfig:limits_twotag}}\\
  \subfloat[]{\includegraphics[width=0.49\textwidth]{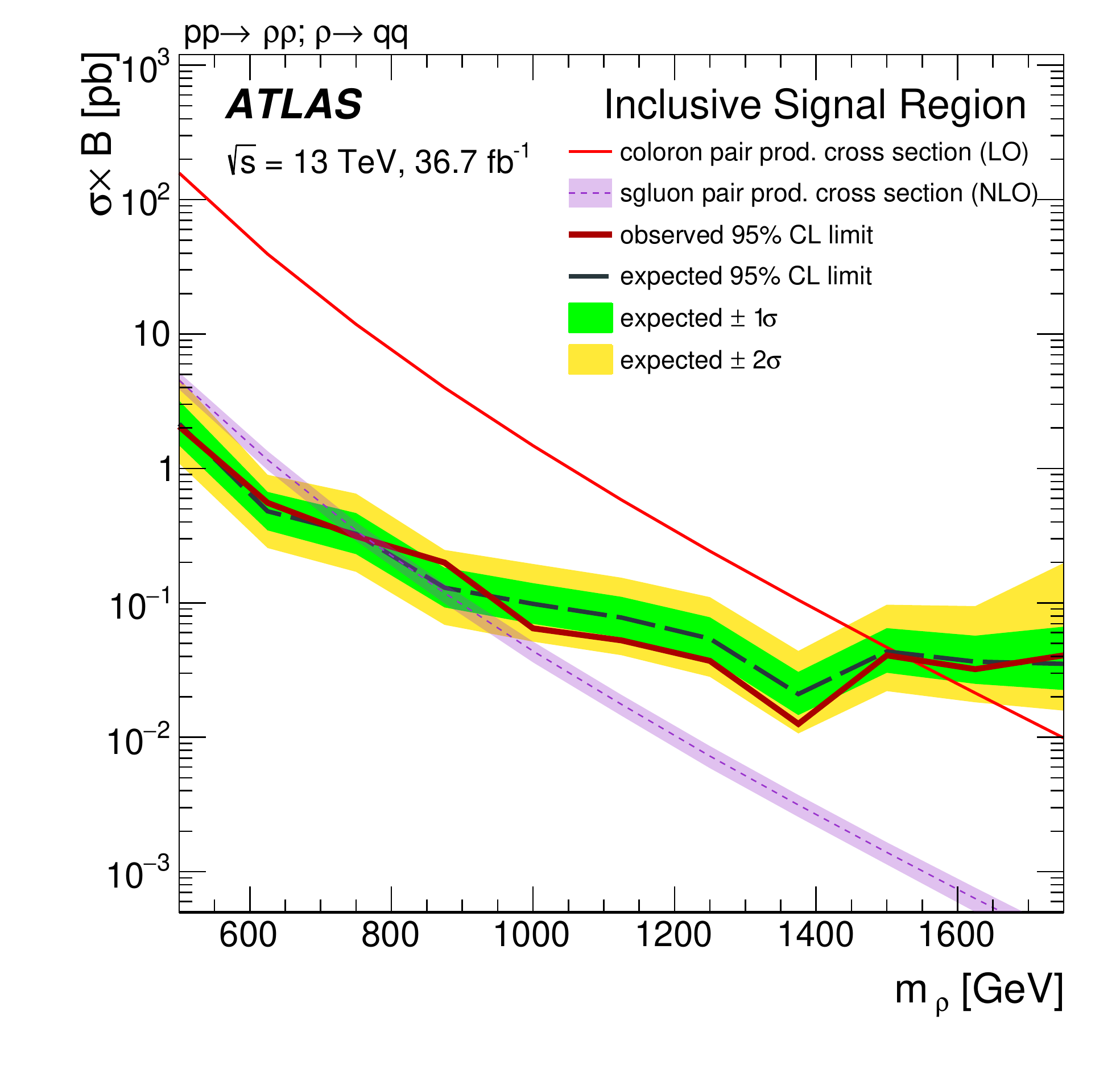}\label{subfig:limits_coloron}}\\
  \caption{\label{fig:limits} The 95\% CL upper limit on the $\sigma\times B$ value 
    compared to the theoretical cross-section for the direct pair-production
    of top squarks with decays into (a) $\bar{q}\bar{q'}$ or (b) $\bar{b}\bar{s}$  and 
    (c) high-mass  colorons decaying into $qq$ and sgluons decaying into $gg$.
    The dashed black and solid red lines show the 95\% CL expected and observed limits respectively,
    including all uncertainties except the theoretical signal cross-section uncertainty.
    The solid green (yellow) band around the expected limit shows the associated $\pm1\sigma$ ($\pm2\sigma$) ranges.
    The shaded coloured cross-section band indicates the $\pm1\sigma$ variations due to theoretical uncertainties
    in the signal production cross-section given by renormalisation and factorisation scale and PDF uncertainties.
    The  region of low top squark mass not shown in the  plot is excluded by Refs.~\cite{Heister:2002jc,Aaltonen:2013hya}.
  }
\end{figure}

\FloatBarrier
\section{Conclusion}
\label{sec:conclusion}
A search is presented for the pair production of coloured resonances, each decaying into two jets.
The analysis uses 36.7~\ifb{} of $\sqrt{s}$ = 13 TeV $pp$ collision data recorded by the ATLAS experiment at the LHC in 2015 and 2016.
An inclusive selection and a selection with two $b$-tagged jets in the event are defined, 
and counting experiments are performed in windows of the average mass of the two resonance candidates.
No significant deviation from the background prediction is observed.
The results are interpreted in a SUSY simplified model with a top squark as the lightest supersymmetric particle,
which is pair-produced and decays promptly into two quarks through  $R$-parity-violating couplings.
For decays into two quarks, top squark masses in the range $100\,\GeV<m_{\tilde{t}}<410\,\GeV$ are excluded at 95\%~CL. 
If the top squark decays into a $b$-quark and a light quark, masses in the ranges
$100\,\GeV<m_{\tilde{t}}<470\,\GeV$ and $480\,\GeV<m_{\tilde{t}}<610\,\GeV$ are excluded at 95\%~CL.
Limits on the pair production of scalar gluons with decays into two gluons reach masses of 800~\GeV.
Vector colour-octet resonances  coupling only to light quarks are excluded up to masses of 1500~\GeV. 
The results  improve upon previous Run 1 searches and extend the constraints on  top squark masses.
\clearpage

\section*{Acknowledgements}


We thank CERN for the very successful operation of the LHC, as well as the
support staff from our institutions without whom ATLAS could not be
operated efficiently.

We acknowledge the support of ANPCyT, Argentina; YerPhI, Armenia; ARC, Australia; BMWFW and FWF, Austria; ANAS, Azerbaijan; SSTC, Belarus; CNPq and FAPESP, Brazil; NSERC, NRC and CFI, Canada; CERN; CONICYT, Chile; CAS, MOST and NSFC, China; COLCIENCIAS, Colombia; MSMT CR, MPO CR and VSC CR, Czech Republic; DNRF and DNSRC, Denmark; IN2P3-CNRS, CEA-DRF/IRFU, France; SRNSFG, Georgia; BMBF, HGF, and MPG, Germany; GSRT, Greece; RGC, Hong Kong SAR, China; ISF, I-CORE and Benoziyo Center, Israel; INFN, Italy; MEXT and JSPS, Japan; CNRST, Morocco; NWO, Netherlands; RCN, Norway; MNiSW and NCN, Poland; FCT, Portugal; MNE/IFA, Romania; MES of Russia and NRC KI, Russian Federation; JINR; MESTD, Serbia; MSSR, Slovakia; ARRS and MIZ\v{S}, Slovenia; DST/NRF, South Africa; MINECO, Spain; SRC and Wallenberg Foundation, Sweden; SERI, SNSF and Cantons of Bern and Geneva, Switzerland; MOST, Taiwan; TAEK, Turkey; STFC, United Kingdom; DOE and NSF, United States of America. In addition, individual groups and members have received support from BCKDF, the Canada Council, CANARIE, CRC, Compute Canada, FQRNT, and the Ontario Innovation Trust, Canada; EPLANET, ERC, ERDF, FP7, Horizon 2020 and Marie Sk{\l}odowska-Curie Actions, European Union; Investissements d'Avenir Labex and Idex, ANR, R{\'e}gion Auvergne and Fondation Partager le Savoir, France; DFG and AvH Foundation, Germany; Herakleitos, Thales and Aristeia programmes co-financed by EU-ESF and the Greek NSRF; BSF, GIF and Minerva, Israel; BRF, Norway; CERCA Programme Generalitat de Catalunya, Generalitat Valenciana, Spain; the Royal Society and Leverhulme Trust, United Kingdom.

The crucial computing support from all WLCG partners is acknowledged gratefully, in particular from CERN, the ATLAS Tier-1 facilities at TRIUMF (Canada), NDGF (Denmark, Norway, Sweden), CC-IN2P3 (France), KIT/GridKA (Germany), INFN-CNAF (Italy), NL-T1 (Netherlands), PIC (Spain), ASGC (Taiwan), RAL (UK) and BNL (USA), the Tier-2 facilities worldwide and large non-WLCG resource providers. Major contributors of computing resources are listed in Ref.~\cite{ATL-GEN-PUB-2016-002}.


\printbibliography

\newpage
\begin{flushleft}
{\Large The ATLAS Collaboration}

\bigskip

M.~Aaboud$^\textrm{\scriptsize 137d}$,
G.~Aad$^\textrm{\scriptsize 88}$,
B.~Abbott$^\textrm{\scriptsize 115}$,
O.~Abdinov$^\textrm{\scriptsize 12}$$^{,*}$,
B.~Abeloos$^\textrm{\scriptsize 119}$,
S.H.~Abidi$^\textrm{\scriptsize 161}$,
O.S.~AbouZeid$^\textrm{\scriptsize 139}$,
N.L.~Abraham$^\textrm{\scriptsize 151}$,
H.~Abramowicz$^\textrm{\scriptsize 155}$,
H.~Abreu$^\textrm{\scriptsize 154}$
% ...
\bigskip
\\
% ...
$^{12}$ Institute of Physics, Azerbaijan Academy of Sciences, Baku, Azerbaijan\\
% ...
$^{88}$ CPPM, Aix-Marseille Universit{\'e} and CNRS/IN2P3, Marseille, France\\
% ...
$^{115}$ Homer L. Dodge Department of Physics and Astronomy, University of Oklahoma, Norman OK, United States of America\\
% ...
$^{119}$ LAL, Univ. Paris-Sud, CNRS/IN2P3, Universit{\'e} Paris-Saclay, Orsay, France\\
% ...
$^{137}$ $^{(a)}$ Facult{\'e} des Sciences Ain Chock, R{\'e}seau Universitaire de Physique des Hautes Energies - Universit{\'e} Hassan II, Casablanca; $^{(b)}$ Centre National de l'Energie des Sciences Techniques Nucleaires, Rabat; $^{(c)}$ Facult{\'e} des Sciences Semlalia, Universit{\'e} Cadi Ayyad, LPHEA-Marrakech; $^{(d)}$ Facult{\'e} des Sciences, Universit{\'e} Mohamed Premier and LPTPM, Oujda; $^{(e)}$ Facult{\'e} des sciences, Universit{\'e} Mohammed V, Rabat, Morocco\\
% ...
$^{139}$ Santa Cruz Institute for Particle Physics, University of California Santa Cruz, Santa Cruz CA, United States of America\\
% ...
$^{151}$ Department of Physics and Astronomy, University of Sussex, Brighton, United Kingdom\\
% ...
$^{154}$ Department of Physics, Technion: Israel Institute of Technology, Haifa, Israel\\
$^{155}$ Raymond and Beverly Sackler School of Physics and Astronomy, Tel Aviv University, Tel Aviv, Israel\\
% ...
$^{161}$ Department of Physics, University of Toronto, Toronto ON, Canada\\
% ...
$^{a}$ Also at Department of Physics, King's College London, London, United Kingdom\\
$^{b}$ Also at Institute of Physics, Azerbaijan Academy of Sciences, Baku, Azerbaijan\\
$^{c}$ Also at Novosibirsk State University, Novosibirsk, Russia\\
$^{d}$ Also at TRIUMF, Vancouver BC, Canada\\
$^{e}$ Also at Department of Physics {\&} Astronomy, University of Louisville, Louisville, KY, United States of America\\
% ...
$^{*}$ Deceased
\end{flushleft}


\end{document}